\documentclass[twocolumn]{aastex631}

\usepackage{amsmath}
\usepackage{amssymb}
\usepackage{bm}

\newcommand{\TRC}{\mathrm{TRC}}

\shorttitle{Source-local DM in Young FRB Environments}
\shortauthors{Fang et al.}

\begin{document}

\title{Propagation Diagnostics of Supernova Remnant Environments around Young Repeating Fast Radio Bursts. I. Hydrodynamic Evolution of the Source-local Dispersion Measure}

\author[0009-0001-8519-0039]{Xiang-Er Fang}
\affiliation{School of Optoelectronic and Communication Engineering, Xiamen University of Technology, Xiamen 361024, People's Republic of China}

\author[0000-0002-2171-9861]{Zhen-Yin Zhao}
\affiliation{School of Astronomy and Space Science, Nanjing University, Nanjing 210023, People's Republic of China}

\author{Yi-Qing Lin}
\affiliation{School of Optoelectronic and Communication Engineering, Xiamen University of Technology, Xiamen 361024, People's Republic of China}

\author{Hao-Tian Lan}
\affiliation{School of Astronomy and Space Science, Nanjing University, Nanjing 210023, People's Republic of China}

\author[0000-0003-4157-7714]{Fa-Yin Wang}
\affiliation{School of Astronomy and Space Science, Nanjing University, Nanjing 210023, People's Republic of China}
\email{fayingwang@nju.edu.cn}
\affiliation{Key Laboratory of Modern Astronomy and Astrophysics (Nanjing University), Ministry of Education, Nanjing 210093, People's Republic of China}

\correspondingauthor{Xiang-Er Fang}
\email{xefang@xmut.edu.cn}

\begin{abstract}

Repeating fast radio bursts may reside in young supernova remnant (SNR) environments whose evolving plasma contributes to the observed dispersion measure (DM). We use two-dimensional axisymmetric hydrodynamic simulations to study the interaction between a continuous anisotropic wind from a young neutron star and homologously expanding supernova ejecta. We follow the evolution to approximately \(160\ \mathrm{yr}\) and calculate the source-local DM along different viewing directions, using a passive tracer to separate wind and nonwind contributions. In the fiducial model, the strongly polar-focused wind inflates a low-density cavity, while the swept-up shell remains broadly rounded and the DM shows moderate angular variation. The DM is dominated by ejecta and swept-up nonwind material. The solid-angle-averaged ambient-subtracted excess DM declines throughout the evolution, approximately following \(t_{\rm age}^{-2}\) during the first several tens of years and becoming modestly steeper later. Variations in wind and ejecta parameters modify the normalization, early evolution, and viewing-angle dependence, but the angle-averaged DM declines in all models, while different bipolar wind profiles produce similar long-term evolution. For FRB 20190520B, the fiducial model reaches a decline rate comparable to the source-frame value inferred from observations at approximately \(20\ \mathrm{yr}\), when the mean excess DM is approximately \(1.8\times10^{2}\ \mathrm{pc\,cm^{-3}}\). Thus, such a young environment can retain a substantial electron column while producing a rapid secular decrease. Repeater diversity suggests that SNR-driven expansion may coexist with additional time-dependent plasma structures or ionization changes.

\end{abstract}


\section{Introduction}
\label{sec:introduction}

Fast radio bursts (FRBs) are bright, millisecond-duration radio transients whose large dispersion measures indicate an extragalactic or cosmological origin \citep{Lorimer2007,CordesChatterjee2019,Petroff2019}. The detection of repeating bursts from some sources shows that at least part of the FRB population is powered by long-lived central engines. Several well-studied repeaters are also associated with compact persistent radio sources \citep{Chatterjee2017,Niu2022,Bruni2024,Bruni2025,Ibik2024,Moroianu2026}, large or evolving dispersion measures (DMs; \citealt{Hessels2019,Oostrum2020,Li2021,Wang2025,Snelders2025,Niu2022,Niu2026,Moroianu2026,Cook2026,Cui2026}), or extreme or time-variable rotation measures (RMs; \citealt{Michilli2018,Hilmarsson2021,Xu2022,Anna-Thomas2023,Mckinven2023,Mckinven2023b,Feng2025,Ng2025,Li2026}). These properties suggest that some repeating FRBs reside in dense, magnetized, and dynamically evolving environments \citep{WangEtAl2022}. Propagation observables therefore provide an important probe of plasma surrounding FRB sources on scales that are generally too compact to resolve directly.

The DM is the integrated free-electron column density along the line of sight, \({\rm DM}=\int n_e\,dl\). The observed value contains contributions from the Milky Way, the Galactic halo, the intergalactic medium, the host galaxy, and the immediate source environment. Separating these components is difficult because the host and source-local contributions are not measured independently. Temporal evolution provides an additional constraint: the Milky Way and intergalactic terms are expected to remain nearly constant over typical monitoring intervals, whereas compact host or source-local plasma can evolve on observable timescales \citep{YangZhang2017,PiroGaensler2018,ZhaoWang2021,Zhao2021,WangEtAl2022,Zhao2023,WFY2025}. A secular DM change can therefore reveal an evolving source-associated electron column and may constrain its density, expansion history, ionization state, and characteristic evolutionary timescale, although these properties cannot be inferred uniquely from DM alone.

FRB 20190520B provides a clear example of a rapid and persistent secular decrease. It is associated with a compact persistent radio source in a dwarf host galaxy and has a large, although uncertain, host-associated DM contribution \citep{Niu2022}. Long-term FAST monitoring shows an approximately monotonic observer-frame decline at a rate of about \(10\ \mathrm{pc\,cm^{-3}\,yr^{-1}}\) over about four years \citep{WFY2025,Niu2026}. FRB 20220529A evolves more slowly, with an observer-frame decline rate of \(-0.881\pm0.001\ {\rm pc\,cm^{-3}\,yr^{-1}}\) \citep{Pandhi2026}. FRB 20121102A has a more complex history in which the DM first increased and later declined \citep{Hessels2019,Oostrum2020,Li2021,Wang2025,Snelders2025}. Its late-time decrease may contain an expansion-driven component, whereas the full nonmonotonic evolution likely requires additional time-dependent plasma structures or changes in the ionization state \citep{WFY2025,Wang2025}. These sources motivate models that can explain a robust secular decline while allowing more complex source-local evolution.

A young compact object embedded in supernova ejecta provides a natural framework for such environments. A newly formed neutron star, such as a magnetar, can continue to inject energy and mass through a central wind or episodic giant flares \citep{Kashiyama2016,Murase2016,MargalitMetzger2018,ZhaoWang2021,Rahaman2025,Zhao2026}. In this work, we focus on continuous energy injection by the central-engine wind. The wind inflates a nebula, drives shocks, and compresses the surrounding ejecta into a shell. For freely expanding material with an approximately fixed ionization fraction, \(n_e\propto t_{\rm age}^{-3}\) while the characteristic path length scales as \(l\propto t_{\rm age}\), giving \({\rm DM}\propto t_{\rm age}^{-2}\). Analytic and one-zone studies have examined the DM, RM, ionization state, and radio transparency of young supernova remnants (SNRs) and magnetar wind nebulae \citep{Piro2016,YangZhang2017,PiroGaensler2018,MargalitMetzger2018,ZhaoWang2021}. These calculations show that shocks, photoionization, recombination, and the ambient medium can modify the simplest freely expanding scaling.

More recently, \citet{ZhangEtAl2026SNR} modeled single-star and binary-stripped progenitors with one-dimensional SNR simulations. Their calculations include nonequilibrium ionization and radiative cooling in the shocked plasma, while treating the ionization of the unshocked ejecta and circumstellar material parametrically. They found that the shocked region generally contributes only \(\lesssim10\ \mathrm{pc\,cm^{-3}}\), whereas the unshocked ejecta supply most of the time-varying DM, with early-time decline indices of approximately \(1.8\)--\(1.9\). These models provide a detailed treatment of progenitor structure, composition, ionization, and cooling in spherical symmetry.

A multidimensional wind--ejecta interaction raises additional questions that cannot be addressed in spherical symmetry. An anisotropic central wind can excavate a low-density cavity, redistribute the ejecta, and produce different electron columns along different viewing directions. It is not obvious whether the source-local DM should be dominated by the directly injected wind, the expanding ejecta, or the wind-compressed shell, nor how strongly the angular structure of the wind should be imprinted on the shell and the observed DM. The dependence of the DM normalization, angular spread, and long-term decline on the source and ejecta parameters also remains uncertain.

In this work, we use two-dimensional axisymmetric hydrodynamic simulations performed with the PLUTO code \citep{Mignone2007} to study a continuous anisotropic wind interacting with homologously expanding supernova ejecta. We follow the wind nebula and shell evolution, calculate the source-local DM along different viewing directions, and use a passive tracer to separate wind and nonwind contributions. Controlled model variations and long-duration calculations are used to test the dependence on the wind, ejecta, initialization, and angular injection profile. We use FRB 20190520B as the main quantitative observational comparison, discuss FRB 20220529A and FRB 20121102A as complementary cases, and compare our characteristic ages and electron columns with the one-dimensional results of \citet{ZhangEtAl2026SNR}. Our goal is to identify the physical origin and generic evolution of the source-local DM rather than to fit an individual repeater in detail.

The simulations show that the wind inflates a low-density cavity and redistributes the ejecta into a compressed shell. Despite the strong polar concentration of the injected power, the outer shell remains broadly rounded and the angular variation of the source-local DM is much weaker than the angular contrast of the injected wind power. The polar sight line nevertheless retains the clearest directional signature, with a larger wind fraction and a less smooth temporal evolution. The directly injected wind contributes only a small fraction of the electron column; the DM is dominated by ejecta-associated nonwind material. The solid-angle-averaged ambient-subtracted excess DM declines in every model, approximately as \(t_{\rm age}^{-2}\) during the first several tens of years and somewhat more steeply later. Model parameters affect the normalization, early evolution, and angular spread, but the long-term evolution is only weakly sensitive to the adopted bipolar wind profile. For FRB 20190520B, the fiducial model reaches the source-frame decline rate inferred from observations at an age of about \(20\ {\rm yr}\), when the mean excess DM is approximately \(1.8\times10^2\ {\rm pc\,cm^{-3}}\). These results show that global expansion and dilution control the secular decline, while the anisotropic wind determines the morphology and detailed sight line dependence.

Section~\ref{sec:methods} describes the numerical model and DM diagnostics. Sections~\ref{sec:fiducial}--\ref{sec:radial_dm} present the fiducial evolution, parameter dependence, long-term robustness, and radial origin of the electron column. Section~\ref{sec:repeating_frb_comparison} compares the simulations with repeating FRB observations, Section~\ref{sec:discussion} discusses the physical interpretation and model limitations, and Section~\ref{sec:summary_conclusions} summarizes the main conclusions.

\section{Numerical Methods}
\label{sec:methods}

\begin{figure*}
\centering
\includegraphics[width=0.92\textwidth]{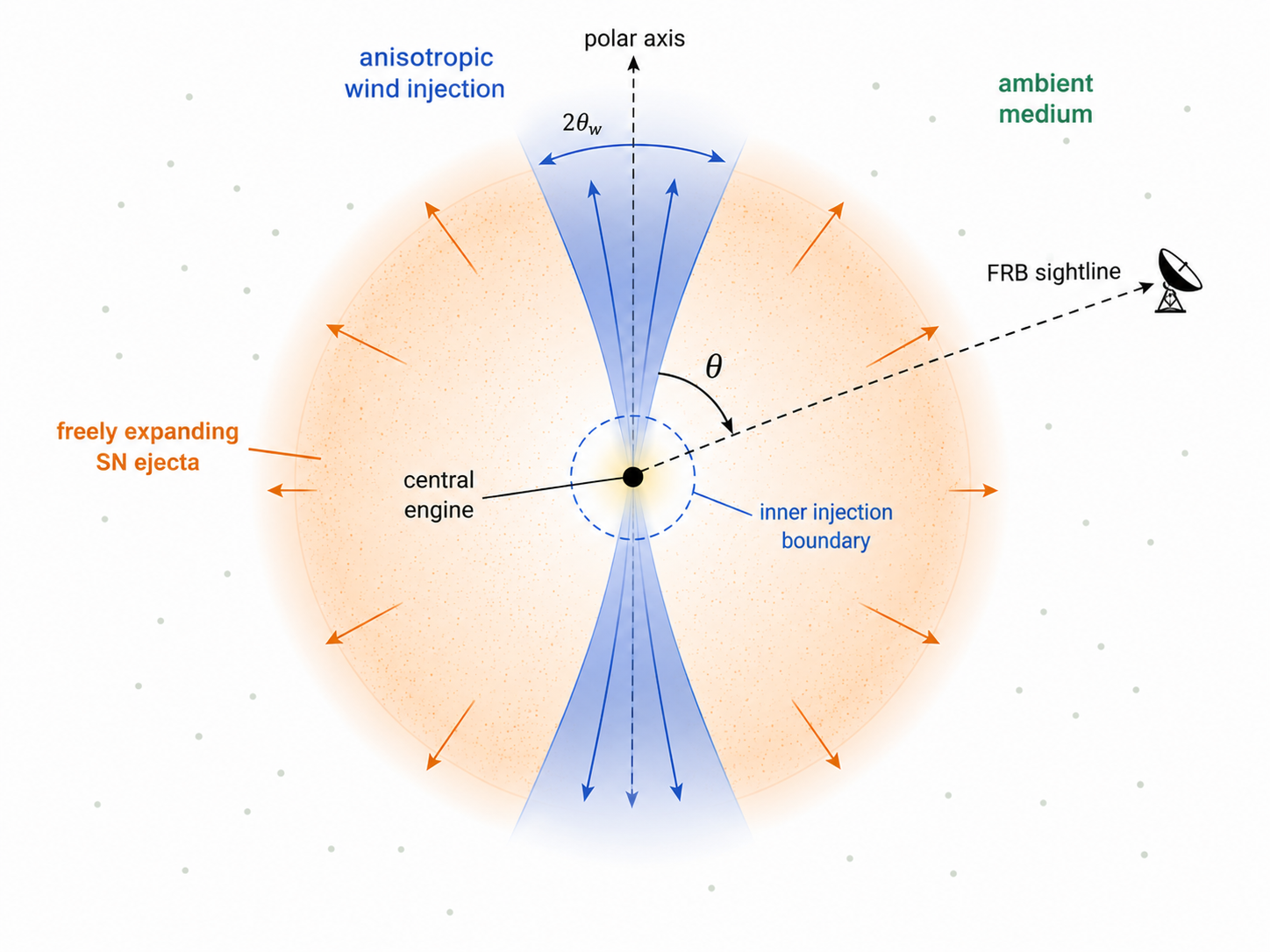}
\caption{
Schematic of the initial wind--ejecta model. A young central engine is embedded in freely expanding supernova ejecta and injects an anisotropic bipolar wind through the inner radial boundary. The ejecta expand into a low-density ambient medium. The dashed line shows a representative FRB sight line at viewing angle $\theta$, measured from the polar axis. The full opening angle of each region of enhanced polar injection is shown as $2\theta_{\rm w}$, where $\theta_{\rm w}\equiv\theta_{\rm cone}$ is the cone half-opening angle used in the numerical prescription below. The blue cones indicate the regions in which the injected wind power is concentrated and should not be interpreted as an already developed wind cavity. The diagram is schematic and not to scale.
}
\label{fig:initial_wind_ejecta_model}
\end{figure*}

Figure~\ref{fig:initial_wind_ejecta_model} summarizes the physical setup at the beginning of the calculation. A young central engine is embedded in freely expanding supernova ejecta, which are surrounded by a low-density ambient medium. An anisotropic bipolar wind, with its power concentrated toward the polar directions, is imposed through the inner radial boundary, and the source-local DM is evaluated along radial sight lines specified by the viewing angle \(\theta\). No wind cavity or compressed shell is imposed in the initial condition; both develop self-consistently through the subsequent wind--ejecta interaction.

\subsection{Basic Equations}
\label{subsec:basic_equations}

We investigate the interaction between a continuous central wind and the surrounding supernova ejecta by solving the ideal hydrodynamic equations. 
The simulations are performed in two-dimensional spherical coordinates $(r,\theta)$, assuming axisymmetry around the polar axis. 
The basic equations can be written in conservative form as
\begin{equation}
\frac{\partial \rho}{\partial t}
+
\nabla\cdot(\rho {\bf v})
=0,
\label{eq:mass}
\end{equation}
\begin{equation}
\frac{\partial (\rho {\bf v})}{\partial t}
+
\nabla\cdot(\rho {\bf v}{\bf v}+P{\bf I})
=0,
\label{eq:momentum}
\end{equation}
and
\begin{equation}
\frac{\partial E}{\partial t}
+
\nabla\cdot\left[(E+P){\bf v}\right]
=0.
\label{eq:energy}
\end{equation}
Here $\rho$, ${\bf v}$, and $P$ are the gas mass density, velocity, and thermal pressure, respectively. 
The total energy density is
\begin{equation}
E=e+\frac{1}{2}\rho v^2,
\label{eq:total_energy}
\end{equation}
where $e$ is the internal energy density. 
We adopt an ideal-gas equation of state,
\begin{equation}
P=(\gamma-1)e,
\label{eq:eos}
\end{equation}
with an adiabatic index $\gamma=5/3$.

Equations~(\ref{eq:mass})--(\ref{eq:energy}) are solved inside the computational domain without explicit volumetric source terms. 
The central wind is instead introduced through time-dependent inner-boundary conditions at $r=r_{\rm in}$, which specify the mass, momentum, and energy fluxes entering the grid. 
Thus, the wind is supplied through the boundary fluxes of the finite-volume domain, rather than injected as source terms within the computational cells.

We do not include self-gravity, external gravity, radiative cooling, or magnetic fields in the present hydrodynamic calculations. 
This treatment allows us to focus on the dynamical interaction between the injected wind and the freely expanding ejecta, and on the resulting evolution of the dispersion measure. 
A self-consistent treatment of magnetic fields, RM evolution, and persistent radio emission is deferred to future work.

In order to distinguish the injected wind material from the original nonwind material in the computational domain, we evolve a passive tracer density, $\rho_{\rm trc}$, according to
\begin{equation}
\frac{\partial \rho_{\rm trc}}{\partial t}
+
\nabla\cdot(\rho_{\rm trc}{\bf v})
=0.
\label{eq:tracer}
\end{equation}
We define the corresponding dimensionless tracer fraction (TRC) as
\begin{equation}
\TRC \equiv \frac{\rho_{\rm trc}}{\rho}.
\label{eq:tracer_fraction}
\end{equation}
The tracer is set to $\TRC=1$ for the gas injected from the inner wind boundary and $\TRC=0$ for the initial ejecta and ambient medium. 
Thus, gas with $\TRC\simeq 1$ is wind dominated, while gas with $\TRC\simeq 0$ is dominated by the original nonwind material. 
This tracer will be used below to separate the wind and nonwind contributions to the dispersion measure.

\subsection{Initial Supernova Ejecta and Ambient Medium}
\label{subsec:initial_ejecta}

As illustrated schematically in Figure~\ref{fig:initial_wind_ejecta_model}, the hydrodynamic calculation is initialized at a time $t_{\rm start}$ after the explosion, when the supernova ejecta are assumed to be in homologous expansion. Inside the ejecta, the radial velocity is prescribed as
\begin{equation}
v_r(r,t_{\rm start})=\frac{r}{t_{\rm start}},
\label{eq:ejecta_velocity}
\end{equation}
and the outer ejecta radius is $R_{\rm ej}=v_{\rm max}t_{\rm start}$, where $v_{\rm max}$ is the maximum ejecta velocity.

For simplicity, we adopt a uniform-density ejecta model. The initial ejecta density is
\begin{equation}
\rho_{\rm ej}(t_{\rm start})
=
\frac{3M_{\rm ej}}{4\pi R_{\rm ej}^3},
\label{eq:ejecta_density}
\end{equation}
for $r\leq R_{\rm ej}$. For a homologously expanding sphere with uniform density, the kinetic energy is $E_{\rm SN}=(3/10)M_{\rm ej}v_{\rm max}^2$, giving
\begin{equation}
v_{\rm max}
=
\left(\frac{10E_{\rm SN}}{3M_{\rm ej}}\right)^{1/2}.
\label{eq:ejecta_vmax}
\end{equation}
In the fiducial model, we adopt $M_{\rm ej}=3.0\,M_\odot$, $E_{\rm SN}=1.0\times10^{51}\ {\rm erg}$, and $t_{\rm start}=1.0\ {\rm yr}$. These parameters determine the initial ejecta radius, density normalization, and homologous velocity field. To examine the effect of a larger ejecta mass, we also perform a High-\(M_{\rm ej}\) model with $M_{\rm ej}=6.0\,M_\odot$ and $E_{\rm SN}=2.0\times10^{51}\ \mathrm{erg}$. This choice preserves the fiducial value of $E_{\rm SN}/M_{\rm ej}$, and hence the same maximum ejecta velocity, while increasing the ejecta density normalization.

Because the computational domain begins at a finite inner radius $r_{\rm in}$, only the ejecta within $r_{\rm in}\leq r\leq R_{\rm ej}$ are evolved explicitly. The exclusion of the innermost ejecta can affect the early-time column density, and the sensitivity to the initialization time is examined using a model with $t_{\rm start}=2.0\ {\rm yr}$.

Outside the ejecta, we place a uniform, low-density ambient medium with $\rho_{\rm amb}=1.0\times10^{-24}\ {\rm g\,cm^{-3}}$ and $v_r=0$. The ejecta and ambient medium are initialized with the same small thermal-pressure floor, with the corresponding internal energy density determined by the ideal-gas equation of state, $e=P/(\gamma-1)$. This initial pressure is dynamically negligible compared with the ejecta kinetic energy and the subsequently injected wind. The ambient medium provides an external background into which the ejecta expand. Both the ejecta and the ambient medium are initialized with $\TRC=0$.

\subsection{Wind Injection Model}
\label{subsec:wind_injection}

The central engine shown in Figure~\ref{fig:initial_wind_ejecta_model} is represented by a continuous, time-dependent wind injected through the inner radial boundary. Motivated by commonly used spin-down-like prescriptions \citep[e.g.,][]{Kashiyama2016,Murase2016,MargalitMetzger2018}, we adopt the generalized wind-luminosity evolution
\begin{equation}
L_{\rm w}(t_{\rm age})
=
L_0\left(1+\frac{t_{\rm age}}{t_0}\right)^{-\alpha},
\label{eq:wind_luminosity}
\end{equation}
where $L_0$ is the luminosity normalization, $t_0$ is a characteristic decay timescale, and $\alpha$ is the decay index. Here $t_{\rm age}$ denotes the physical age of the source. Because the hydrodynamic calculation begins at $t_{\rm start}$, the source age is $t_{\rm age}=t_{\rm start}+t$, where $t$ is the simulation time measured from the beginning of the calculation. For the fiducial model, we adopt $L_0=3.0\times10^{42}\ \mathrm{erg\,s^{-1}}$, $t_0=0.3\ \mathrm{yr}$, and $\alpha=1.3$. We additionally perform a High-\(L_0\) model in which $L_0$ is increased to $9.0\times10^{42}\ \mathrm{erg\,s^{-1}}$, while all other parameters remain at their fiducial values.

The normalization $L_0$ is not the instantaneous wind luminosity at the beginning of the hydrodynamic calculation. For the fiducial start time, $t_{\rm start}=1.0\ \mathrm{yr}$, the initial injected luminosity is
\begin{equation}
L_{\rm w}(t_{\rm start})
=
L_0\left(1+\frac{t_{\rm start}}{t_0}\right)^{-\alpha}
\simeq
4.5\times10^{41}\ \mathrm{erg\,s^{-1}}.
\label{eq:fid_lw_tstart}
\end{equation}
For comparison, the late-time luminosity of a standard magnetic-dipole spin-down model can be estimated to order of magnitude as \citep[e.g.,][]{Kashiyama2016}
\begin{equation}
L_{\rm sd}(t)
\sim
7\times10^{41}
\left(\frac{B_{\rm dip}}{10^{14}\ \mathrm{G}}\right)^{-2}
\left(\frac{t}{1\ \mathrm{yr}}\right)^{-2}
\mathrm{erg\,s^{-1}},
\label{eq:magnetar_lsd_order}
\end{equation}
up to factors of order unity that depend on the spin-down prescription and magnetic geometry. The fiducial value of $L_{\rm w}(t_{\rm start})$ is therefore comparable to the late-time power of a magnetar-strength engine with $B_{\rm dip}$ of order $10^{14}\ \mathrm{G}$. This comparison is used only to motivate the adopted luminosity scale; the injected wind is not tied to a unique magnetic-dipole model.

The injected wind should not be interpreted as the pristine relativistic wind launched directly by the compact object. Instead, it represents an effective large-scale, mass-loaded, and partially thermalized outflow after unresolved wind launching, dissipation, and possible mass loading on scales interior to the computational domain. In particular, the present hydrodynamic prescription does not explicitly follow the Lorentz factor, magnetization, or electron--positron pair multiplicity of a pristine magnetar wind. Such relativistic pair winds can have lepton columns and radio-propagation properties that depend on the pair loading and wind dynamics \citep{CaoEtAl2017,DaiEtAl2017}. Our prescription is therefore intended to characterize the large-scale dynamical coupling between the central-engine outflow and the supernova ejecta rather than the microphysical composition of the inner wind. The luminosity supplied to the resolved wind--ejecta interaction may therefore decline more slowly than the idealized late-time magnetic-dipole scaling, $L_{\rm sd}\propto t^{-2}$. We adopt $\alpha=1.3$ as the fiducial effective wind prescription. To test the sensitivity to this assumption, we also perform an Alpha2 model with $\alpha=2.0$, in which $L_0$ is adjusted so that $L_{\rm w}(t_{\rm start})$ is the same as in the fiducial model.

The wind is anisotropic in the fiducial model. We write the wind energy flux imposed at the inner boundary as
\begin{equation}
F_{\rm w}(\theta,t_{\rm age})
=
\frac{L_{\rm w}(t_{\rm age})}{4\pi r_{\rm in}^2}
A(\theta),
\label{eq:wind_flux}
\end{equation}
where $r_{\rm in}$ is the inner boundary radius and $A(\theta)$ describes the angular distribution of the injected wind power. The angular function is normalized to conserve the total luminosity, such that $(4\pi)^{-1}\int A(\theta)\,d\Omega=1$. We write $A(\theta)=f(\theta)/\langle f\rangle_\Omega$, where $\langle f\rangle_\Omega=(1/2)\int_0^\pi f(\theta)\sin\theta\,d\theta$. The unnormalized fiducial profile is
\begin{equation}
\begin{aligned}
f(\theta) =\;&
f_{\rm floor}
+\frac{1}{2}\left[
1-\tanh\left(\frac{\theta-\theta_{\rm cone}}{\Delta\theta}\right)
\right] \\
&+\frac{1}{2}\left[
1-\tanh\left(\frac{\pi-\theta-\theta_{\rm cone}}{\Delta\theta}\right)
\right].
\end{aligned}
\label{eq:fiducial_wind_angular_profile}
\end{equation}
This form concentrates most of the injected power toward the two polar directions while retaining a finite wind flux at all polar angles. The parameter $f_{\rm floor}$ sets the isotropic floor, $\theta_{\rm cone}$ is the half-opening angle of each polar cone, and $\Delta\theta$ determines the smoothing width at the cone boundary. In the fiducial model, we adopt $f_{\rm floor}=0.1$, $\theta_{\rm cone}=20^\circ$, and $\Delta\theta=4^\circ$. The Narrow-cone model adopts $\theta_{\rm cone}=10^\circ$ to examine the effect of stronger collimation, with all other parameters fixed at their fiducial values. For the long-term robustness test, we additionally consider an alternative smooth-bipolar wind profile described in Appendix~\ref{app:smooth_bipolar_wind}.

At $r=r_{\rm in}$, the wind is injected radially with a fixed effective velocity, $v_r=v_{\rm w}$. Consistent with the effective-outflow interpretation above, the adopted values of $v_{\rm w}$ are not intended to represent the terminal speed of a pristine relativistic wind. We adopt $v_{\rm w}=3.0\times10^9\ \mathrm{cm\,s^{-1}}$ in the fiducial model and examine the sensitivity to this choice using the Low-wind model, for which $v_{\rm w}=1.5\times10^9\ \mathrm{cm\,s^{-1}}$. All other parameters in the Low-wind model are kept at their fiducial values. For a given energy flux $F_{\rm w}$, the injected wind power is divided into kinetic and thermal components, with $\eta_{\rm kin}+\eta_{\rm th}=1$. The wind density and pressure at the inner boundary are determined by
\begin{equation}
\begin{aligned}
\frac{1}{2}\rho_{\rm w}v_{\rm w}^3
&=
\eta_{\rm kin}F_{\rm w}, \\
\frac{\gamma}{\gamma-1}P_{\rm w}v_{\rm w}
&=
\eta_{\rm th}F_{\rm w}.
\end{aligned}
\label{eq:wind_boundary_fluxes}
\end{equation}
The first relation specifies the kinetic-energy flux, whereas the second specifies the thermal component through the enthalpy flux. We adopt $\eta_{\rm kin}=0.7$ and $\eta_{\rm th}=0.3$, so that most of the injected power is initially carried by kinetic energy while a finite fraction is thermalized at injection.

\subsection{Numerical Setup and Model Suite}
\label{subsec:numerical_setup}

We solve the axisymmetric hydrodynamic equations in spherical coordinates $(r,\theta)$ with the PLUTO code. The computational domain extends from $r_{\rm in}=2.0\times10^{16}\ \mathrm{cm}$ to $r_{\rm out}=1.0\times10^{19}\ \mathrm{cm}$ in the radial direction, corresponding to approximately $0.006$--$3.2\ \mathrm{pc}$, and covers $0\leq\theta\leq\pi$ in the angular direction. We adopt a logarithmically spaced radial grid and a uniformly spaced angular grid, with $(N_r,N_\theta)=(600,192)$ active zones. A lower-resolution calculation with $(N_r,N_\theta)=(400,128)$ yields the same qualitative DM evolution and material decomposition, indicating that the main conclusions are not sensitive to the adopted grid resolution. At the inner radial boundary, we impose the time-dependent wind described in Section~\ref{subsec:wind_injection}. An outflow condition is applied at the outer radial boundary, while axisymmetric boundary conditions are imposed at the two polar boundaries. The fiducial parameters are summarized in Table~\ref{tab:fiducial_parameters}; they define a representative young wind--ejecta system rather than a unique model for any individual repeating FRB.

\begin{deluxetable}{ll}
\tabletypesize{\footnotesize}
\tablecaption{Fiducial model parameters\label{tab:fiducial_parameters}}
\tablehead{
\colhead{Quantity} &
\colhead{Fiducial value}
}
\startdata
Wind-luminosity normalization, $L_0$ & $3.0\times10^{42}\ \mathrm{erg\,s^{-1}}$ \\
Wind decay timescale, $t_0$ & $0.3\ \mathrm{yr}$ \\
Wind decay index, $\alpha$ & $1.3$ \\
Wind luminosity at $t_{\rm start}$, $L_{\rm w}(t_{\rm start})$ & $4.5\times10^{41}\ \mathrm{erg\,s^{-1}}$ \\
Ejecta mass, $M_{\rm ej}$ & $3.0\,M_\odot$ \\
Explosion energy, $E_{\rm SN}$ & $1.0\times10^{51}\ \mathrm{erg}$ \\
Simulation start age, $t_{\rm start}$ & $1.0\ \mathrm{yr}$ \\
Effective wind velocity, $v_{\rm w}$ & $3.0\times10^9\ \mathrm{cm\,s^{-1}}$ \\
Kinetic fraction, $\eta_{\rm kin}$ & $0.7$ \\
Thermal fraction, $\eta_{\rm th}$ & $0.3$ \\
Isotropic floor, $f_{\rm floor}$ & $0.1$ \\
Cone half-opening angle, $\theta_{\rm cone}$ & $20^\circ$ \\
Angular smoothing width, $\Delta\theta$ & $4^\circ$ \\
Ambient density, $\rho_{\rm amb}$ & $1.0\times10^{-24}\ \mathrm{g\,cm^{-3}}$ \\
\enddata
\end{deluxetable}

The physical comparison models are summarized in Table~\ref{tab:model_suite}. Except for the Smooth-bipolar run, each model tests one principal physical variation relative to the fiducial setup. The Smooth-bipolar run instead tests the sensitivity of the long-term evolution to the adopted wind angular profile. All models are analyzed using the same source-local DM diagnostics described in Section~\ref{subsec:dm_calculation}.

\begin{deluxetable*}{llll}
\tabletypesize{\footnotesize}
\tablewidth{0pt}
\tablecaption{Simulation suite\label{tab:model_suite}}
\tablehead{
\colhead{Model} &
\colhead{Modified quantity} &
\colhead{Fiducial value} &
\colhead{Test value}
}
\startdata
Fiducial & -- & -- & -- \\
High-$L_0$ & $L_0$ & $3.0\times10^{42}\ \mathrm{erg\,s^{-1}}$ & $9.0\times10^{42}\ \mathrm{erg\,s^{-1}}$ \\
Narrow-cone & $\theta_{\rm cone}$ & $20^\circ$ & $10^\circ$ \\
Low-wind\tablenotemark{a} & $v_{\rm w}$ & $3.0\times10^9\ \mathrm{cm\,s^{-1}}$ & $1.5\times10^9\ \mathrm{cm\,s^{-1}}$ \\
High-$M_{\rm ej}$\tablenotemark{b} & $M_{\rm ej}$, $E_{\rm SN}$ & $M_{\rm ej}=3.0\,M_\odot,\ E_{\rm SN}=1.0\times10^{51}\ \mathrm{erg}$ & $M_{\rm ej}=6.0\,M_\odot,\ E_{\rm SN}=2.0\times10^{51}\ \mathrm{erg}$ \\
$t_{\rm start}=2\ \mathrm{yr}$ & $t_{\rm start}$ & $1.0\ \mathrm{yr}$ & $2.0\ \mathrm{yr}$ \\
Alpha2\tablenotemark{c} & $\alpha$, $L_0$ & $\alpha=1.3,\ L_0=3.0\times10^{42}\ \mathrm{erg\,s^{-1}}$ & $\alpha=2.0,\ L_0=8.4\times10^{42}\ \mathrm{erg\,s^{-1}}$ \\
Smooth-bipolar\tablenotemark{d} & Wind angular profile & Smoothed bipolar-cone & Smooth-bipolar \\
\enddata
\tablenotetext{a}{Lower effective wind velocity, corresponding to greater mass loading at fixed wind luminosity.}
\tablenotetext{b}{Both $M_{\rm ej}$ and $E_{\rm SN}$ are increased by a factor of two, preserving the fiducial ejecta velocity scale while increasing the ejecta density normalization.}
\tablenotetext{c}{$L_0$ is adjusted so that $L_{\rm w}(t_{\rm start})$ matches the fiducial value.}
\tablenotetext{d}{Alternative wind angular profile used for the long-term robustness test and defined in Appendix~\ref{app:smooth_bipolar_wind}.}
\end{deluxetable*}

\subsection{Source-local DM Diagnostics}
\label{subsec:dm_calculation}

For each simulation snapshot, we estimate the electron number density from the gas density as
\begin{equation}
n_e(r,\theta,t_{\rm age})
=
x_e\frac{\rho(r,\theta,t_{\rm age})}{m_p},
\label{eq:electron_density}
\end{equation}
where $m_p$ is the proton mass and $x_e$ is an effective electron abundance per baryon, which depends on both the elemental composition and ionization state of the gas. We adopt a spatially and temporally constant $x_e=1$, corresponding to a fully ionized hydrogen-equivalent normalization. This choice is not intended to represent a specific supernova ejecta composition, but rather to isolate the hydrodynamically driven evolution of the source-local DM. For comparison, fully ionized He-rich or C/O-rich material has $x_e\simeq0.5$ because of its lower charge-to-mass ratio, while incomplete ionization would further reduce the number of free electrons per unit mass. Within the present prescription, adopting a different constant $x_e$ uniformly for all material would rescale the DM and its absolute variation rate at a fixed age without changing the hydrodynamic evolution, the temporal shape and logarithmic decline slope, or the normalized angular statistics. In realistic ejecta, however, composition and ionization can vary among different regions and evolve with time, in which case the relative electron columns of different components and the detailed DM evolution may also change. The viewing angle $\theta$ is measured from the polar axis, as illustrated by the representative sight line in Figure~\ref{fig:initial_wind_ejecta_model}. For a radial line of sight at polar angle $\theta$, the source-local DM is calculated as
\begin{equation}
{\rm DM}_{\rm loc}(\theta,t_{\rm age})
=
\int_{r_{\rm in}}^{r_{\rm out}}
n_e(r,\theta,t_{\rm age})\,dr.
\label{eq:dm_total}
\end{equation}
When evaluating Equation~(\ref{eq:dm_total}), we express the radial path length in parsecs, so that the resulting DM is quoted in units of ${\rm pc\,cm^{-3}}$. The integral covers the simulated source environment; gas inside the inner boundary is not included.

The passive tracer allows us to decompose the source-local DM according to material origin. The wind and nonwind density components are defined as $\rho_{\rm wind}=\TRC\rho$ and $\rho_{\rm nw}=(1-\TRC)\rho$, respectively, with corresponding electron densities $n_{e,\rm wind}=x_e\rho_{\rm wind}/m_p$ and $n_{e,\rm nw}=x_e\rho_{\rm nw}/m_p$. The nonwind component consists of material originating from the initial supernova ejecta and ambient medium, regardless of whether it is subsequently compressed or redistributed by the interaction. Integrating these electron-density components along the same radial path gives ${\rm DM}_{\rm wind}$ and ${\rm DM}_{\rm nw}$, with ${\rm DM}_{\rm loc}={\rm DM}_{\rm wind}+{\rm DM}_{\rm nw}$. When needed, we quote the wind fraction as ${\rm DM}_{\rm wind}/{\rm DM}_{\rm loc}$. We also define the ambient-subtracted excess DM as
\begin{equation}
{\rm DM}_{\rm exc}(\theta,t_{\rm age})
=
{\rm DM}_{\rm loc}(\theta,t_{\rm age})
-
{\rm DM}_{\rm amb},
\label{eq:dm_excess}
\end{equation}
where ${\rm DM}_{\rm amb}$ is the electron column of the unperturbed uniform ambient medium along the same radial path. For the constant ambient density adopted here,
${\rm DM}_{\rm amb}=x_e(\rho_{\rm amb}/m_p)(r_{\rm out}-r_{\rm in})$, with the radial extent expressed in parsecs. This baseline is independent of viewing angle. The excess DM therefore measures the electron column above the unperturbed ambient contribution, including the ejecta, wind material, and any ambient gas redistributed by the interaction.

To characterize the viewing-angle dependence, we compute solid-angle-weighted statistics of the radial DM distribution. The solid-angle-weighted angular mean is
\begin{equation}
\langle {\rm DM}_{\rm loc}\rangle_\Omega
=
\frac{1}{2}
\int_0^\pi
{\rm DM}_{\rm loc}(\theta,t_{\rm age})
\sin\theta\,d\theta.
\label{eq:dm_angle_average}
\end{equation}
The same solid-angle weighting is applied to ${\rm DM}_{\rm exc}$ and to the individual material components when these quantities are averaged over viewing angle. We define $P_{10,\Omega}$ and $P_{90,\Omega}$ as the 10th and 90th percentiles of ${\rm DM}_{\rm loc}(\theta,t_{\rm age})$, using the solid-angle probability weight $dP=(1/2)\sin\theta\,d\theta$. These percentiles enclose the central 80\% of the viewing-angle distribution in solid angle. We quantify the normalized angular DM spread by
\begin{equation}
A_{{\rm DM},\Omega}
=
\frac{P_{90,\Omega}-P_{10,\Omega}}
{\langle {\rm DM}_{\rm loc}\rangle_\Omega}.
\label{eq:angular_dm_spread}
\end{equation}
A larger value of $A_{{\rm DM},\Omega}$ indicates a stronger viewing-angle dependence of the source-local DM.

\section{Fiducial Model Results}
\label{sec:fiducial}

We first examine the fiducial model, focusing on the wind-driven morphology, source-local DM evolution, and viewing-angle dependence produced by the interaction between the anisotropic central wind and the expanding supernova ejecta.

\subsection{Morphological Evolution}
\label{subsec:fiducial_morphology}

\begin{figure*}
\centering
\includegraphics[width=\textwidth]{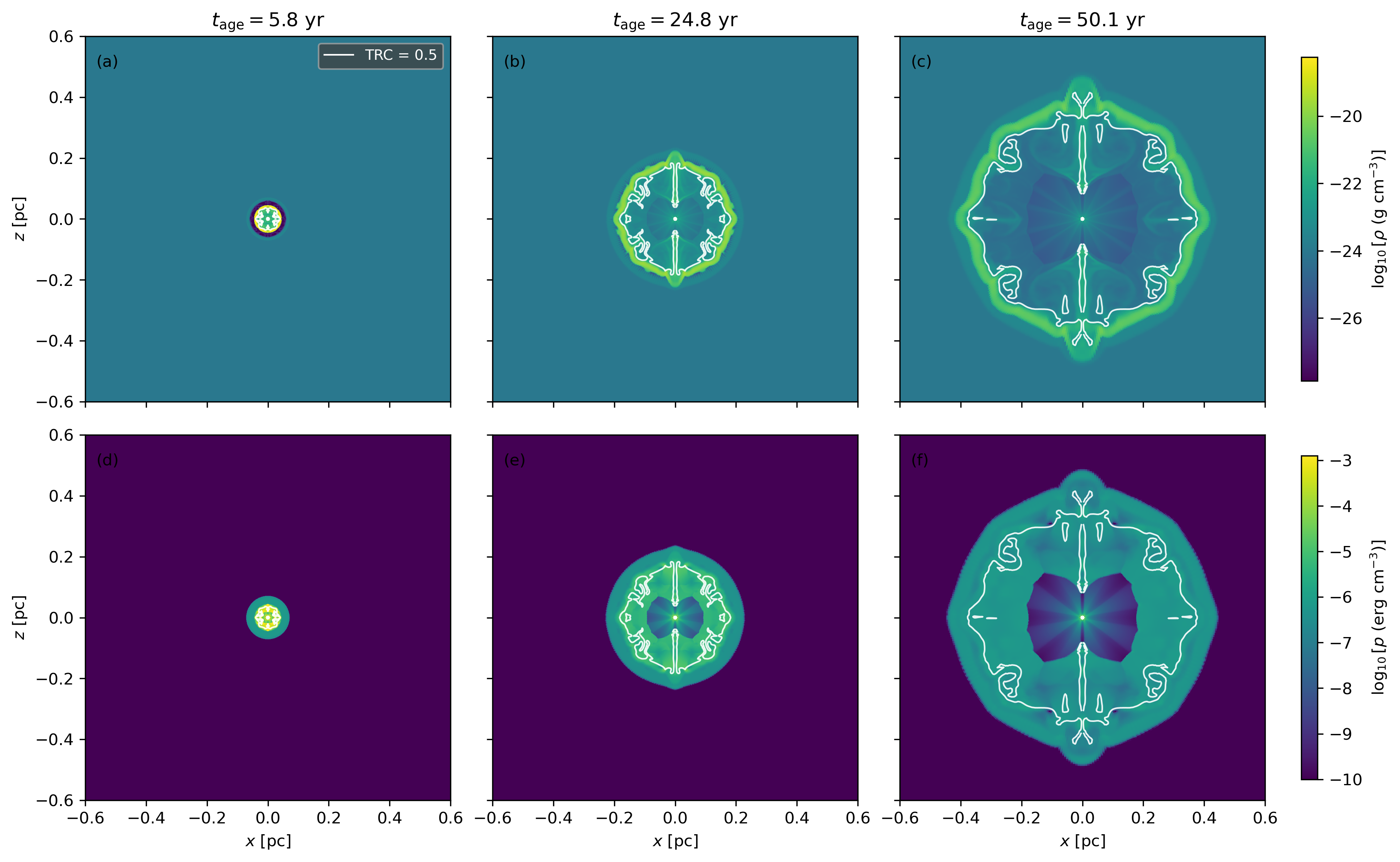}
\caption{
Morphological evolution of the fiducial model at source ages of 5.8, 24.8, and 50.1 yr.
The top and bottom rows show the logarithmic gas density and pressure, respectively.
White contours denote $\TRC=0.5$ and provide a visual guide to the boundary of the wind-rich region.
The bipolar wind injection produces a low-density cavity surrounded by a dense shell whose size and angular structure evolve with source age.
}
\label{fig:fiducial_morphology}
\end{figure*}

Starting from the initial configuration summarized in Figure~\ref{fig:initial_wind_ejecta_model}, the injected wind inflates a nebula and redistributes the surrounding ejecta. Figure~\ref{fig:fiducial_morphology} shows the resulting density and pressure structures of the fiducial model at source ages of 5.8, 24.8, and 50.1 yr. The wind produces a low-density bubble and sweeps the surrounding nonwind material into a dense shell. The overpressured interior drives the shell outward. Because the injected power is concentrated toward the two polar directions, the bubble extends farther along the polar axis than near the equatorial plane. Nevertheless, the outer swept-up shell remains broadly rounded rather than reproducing the strong angular contrast of the injected wind. This morphology suggests that the injection anisotropy is only partly transmitted to the large-scale shell and is consistent with lateral pressure redistribution within the overpressured bubble.

The white contour in Figure~\ref{fig:fiducial_morphology} marks $\TRC=0.5$, which approximately separates wind-dominated material from the original nonwind material. At early times, the wind-inflated region remains compact and deeply embedded within the ejecta. As the system evolves, both the bubble and the surrounding shell expand outward. The characteristic outer extent grows from a few $10^{-2}\ \mathrm{pc}$ at $t_{\rm age}=5.8\ \mathrm{yr}$ to about $0.2\ \mathrm{pc}$ at $24.8\ \mathrm{yr}$ and $0.4$--$0.5\ \mathrm{pc}$ at $50.1\ \mathrm{yr}$. The dense shell remains dominated by nonwind material, indicating that most of the high-density gas surrounding the bubble consists of swept-up and compressed nonwind material rather than newly injected wind.

The flow contains both an inner wind-driven structure and the outer shock system associated with the conventional supernova-remnant evolution \citep{vanDerSwaluwEtAl2001,BlondinEtAl2001}. In the inner region, the injected effective wind is decelerated across a nonspherical reverse/termination-like shock, visible as an inner pressure discontinuity, and subsequently inflates the low-density cavity. The wind-dominated material is then separated from the swept and compressed nonwind material by the wind--ejecta interface traced approximately by the \({\rm TRC}=0.5\) contour. This inner reverse/termination-like shock should not be confused with the conventional SNR reverse shock produced by the interaction of the expanding ejecta with the ambient medium. At larger radii, the ejecta--ambient interaction retains the usual outer blast-wave structure, with the SNR forward shock identifiable as the outermost density and pressure discontinuity. The corresponding ejecta--ambient contact discontinuity and SNR reverse shock are not individually tracked in the present diagnostics because the passive tracer distinguishes wind from nonwind material rather than ejecta from ambient gas. The anisotropic wind primarily distorts the inner cavity, wind--ejecta interface, and compressed nonwind material, while pressure redistribution within the bubble partially smooths these angular differences at larger radii.

\subsection{Source-local DM Evolution and Viewing Angle Dependence}
\label{subsec:fiducial_dm}

\begin{figure*}
\centering
\includegraphics[width=\textwidth]{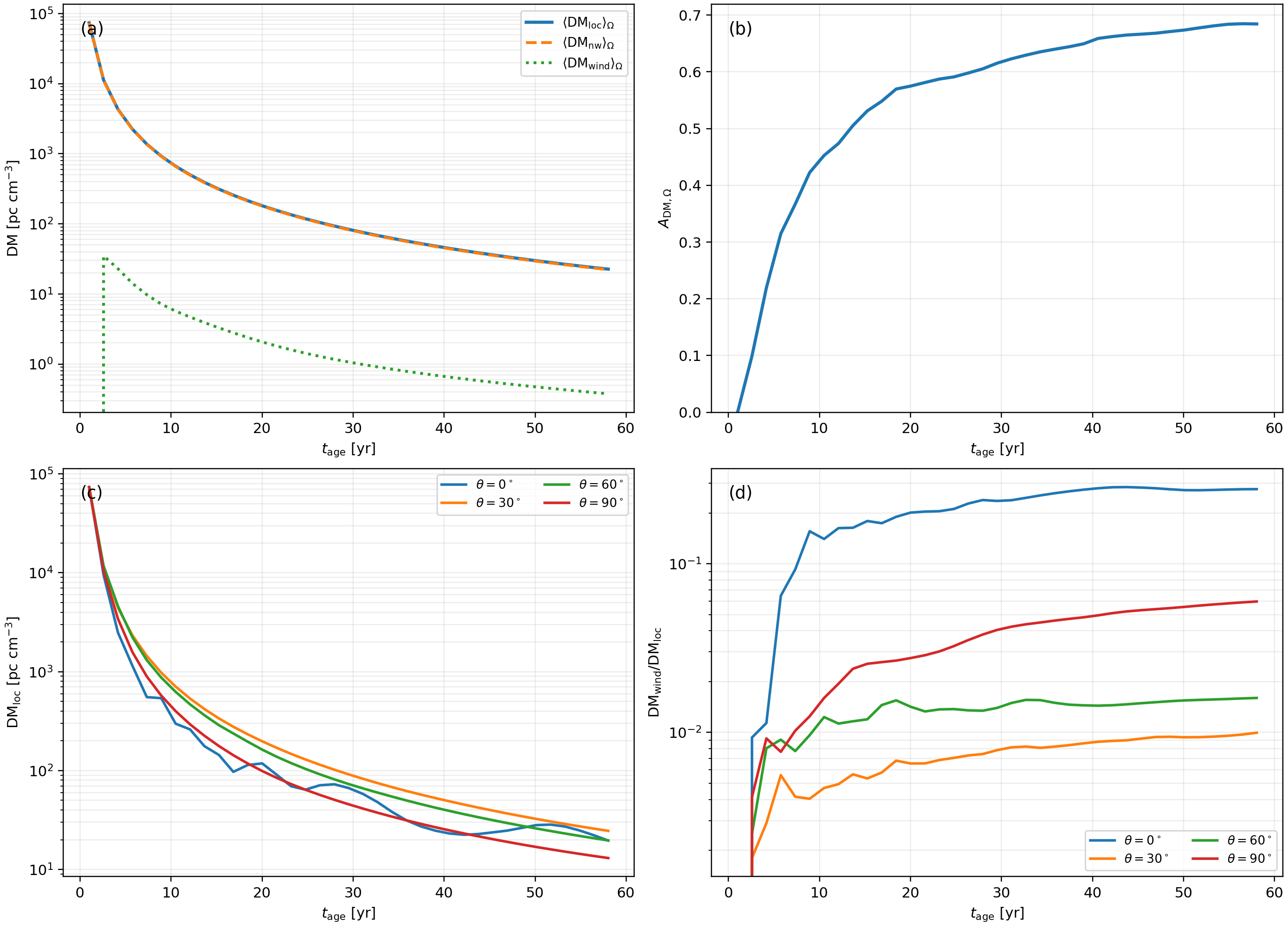}
\caption{
Source-local DM evolution and viewing-angle dependence in the fiducial model.
Panel (a) shows the solid-angle-averaged source-local DM, $\langle{\rm DM}_{\rm loc}\rangle_\Omega$, together with the nonwind and wind contributions, $\langle{\rm DM}_{\rm nw}\rangle_\Omega$ and $\langle{\rm DM}_{\rm wind}\rangle_\Omega$.
Panel (b) shows the normalized angular DM spread, $A_{{\rm DM},\Omega}$.
Panel (c) shows ${\rm DM}_{\rm loc}$ along selected viewing angles.
Panel (d) shows the corresponding wind fraction, ${\rm DM}_{\rm wind}/{\rm DM}_{\rm loc}$.
The mean electron column is controlled by nonwind material, whereas the relative wind contribution is most prominent along the polar sight line.
}
\label{fig:fiducial_dm}
\end{figure*}

Figure~\ref{fig:fiducial_dm} shows the source-local DM evolution in the fiducial model. The solid-angle-averaged source-local DM, $\langle{\rm DM}_{\rm loc}\rangle_\Omega$, decreases rapidly with source age, from $\gtrsim10^4\ \mathrm{pc\,cm^{-3}}$ during the first few years to about $20\ \mathrm{pc\,cm^{-3}}$ by $t_{\rm age}\simeq60\ \mathrm{yr}$. This decline primarily reflects the expansion and dilution of the ejecta and swept-up nonwind material, whose electron column decreases as the system grows. The solid-angle-averaged nonwind contribution closely tracks $\langle{\rm DM}_{\rm loc}\rangle_\Omega$ throughout the evolution, showing that the dominant electron column arises from the swept-up shell and the ejecta outside the wind-inflated bubble. By contrast, $\langle{\rm DM}_{\rm wind}\rangle_\Omega$ remains below a few percent of $\langle{\rm DM}_{\rm loc}\rangle_\Omega$. The wind is therefore dynamically important in shaping the system but contributes little to the angle-averaged electron column because it occupies a relatively low-density cavity.

The viewing-angle dependence is quantified by the normalized angular DM spread \(A_{{\rm DM},\Omega}\) defined in Equation~(\ref{eq:angular_dm_spread}). As shown in Figure~\ref{fig:fiducial_dm}(b), \(A_{{\rm DM},\Omega}\) rises rapidly during the first \(\sim20\ \mathrm{yr}\) and then increases more gradually, approaching a moderate value of approximately \(0.7\) by \(t_{\rm age}\simeq60\ \mathrm{yr}\). Despite the strong polar concentration of the injected wind, the angular DM spread remains moderate, indicating that the injection anisotropy does not translate into an equally strong anisotropy of the electron column. This behavior is consistent with the broadly rounded outer shell and with the dominance of ejecta-associated material in the total DM. The selected sight lines in Figure~\ref{fig:fiducial_dm}(c) all show an overall secular decline, but their detailed evolution and relative ordering vary with time. In particular, the polar sight line exhibits a less smooth evolution, with local changes in slope and brief intervals of flattening or increase. This behavior reflects the stronger influence of the bipolar wind on the polar cavity and on the compressed nonwind material at its boundary. Away from the axis, the DM evolution is comparatively smoother and is controlled mainly by the expansion and dilution of ejecta-associated material. Thus, anisotropic wind dynamics introduce sight line-dependent temporal structure while the solid-angle-averaged DM continues to decline monotonically.

The wind fraction is more anisotropic than the total DM. Figure~\ref{fig:fiducial_dm}(d) shows that \({\rm DM}_{\rm wind}/{\rm DM}_{\rm loc}\) along the polar sight line rises to approximately \(0.2\)--\(0.3\), whereas the other selected sight lines remain below approximately \(0.06\). Even along the polar direction, however, the source-local DM remains dominated by nonwind material because the wind-rich cavity has a substantially lower density than the surrounding ejecta and swept-up shell. The directional imprint of the wind is therefore most apparent in the enhanced polar wind fraction and the less smooth polar DM evolution, rather than as a comparably strong anisotropy in the total electron column.

\section{Parameter Dependence}
\label{sec:parameter_dependence}

\begin{figure*}
\centering
\includegraphics[width=\textwidth]{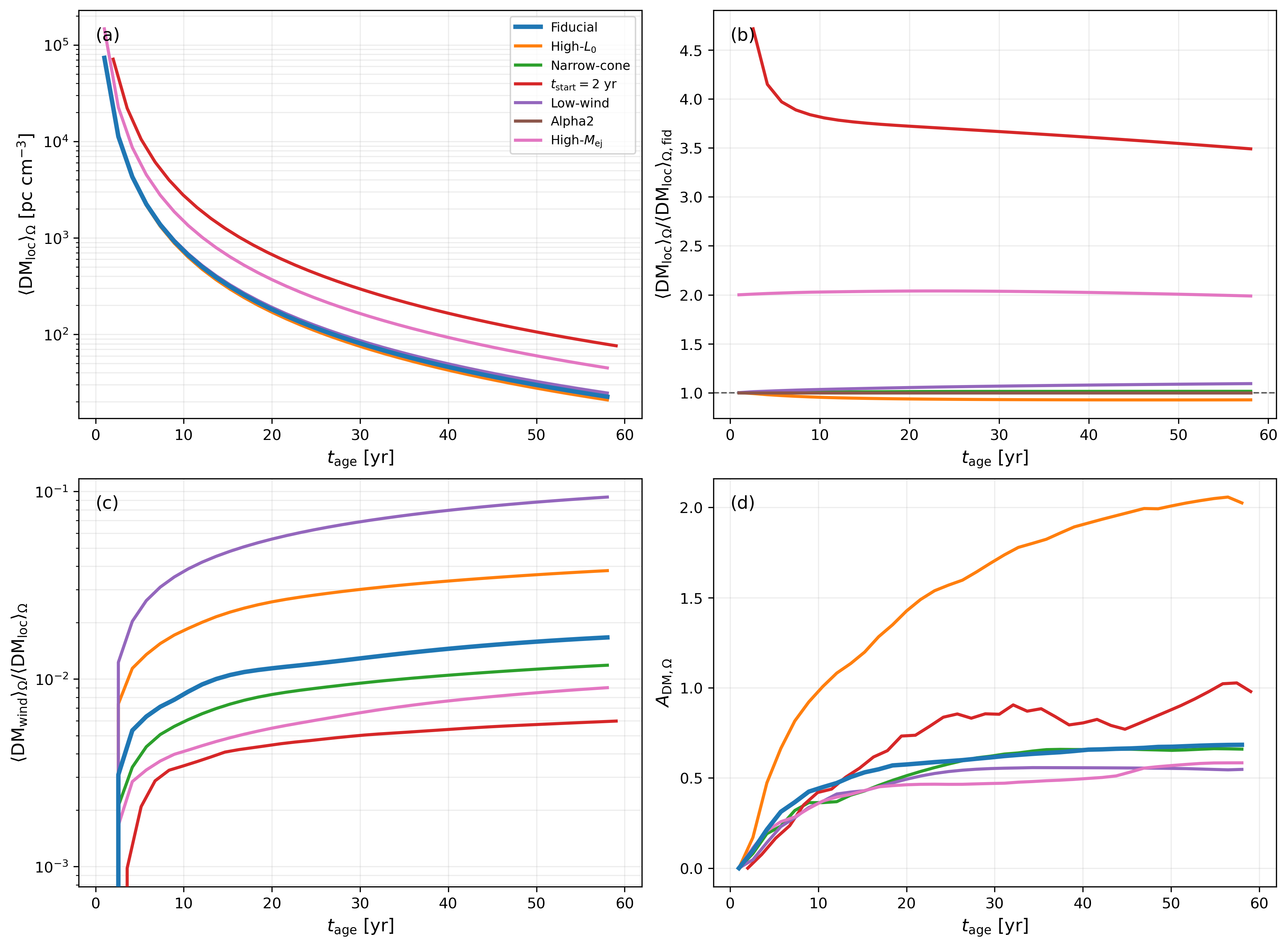}
\caption{
Parameter dependence of the source-local DM evolution.
The models shown are Fiducial, High-$L_0$, Narrow-cone, $t_{\rm start}=2~{\rm yr}$, Low-wind, Alpha2, and High-$M_{\rm ej}$.
Panel (a) shows the solid-angle-averaged source-local DM, $\langle{\rm DM}_{\rm loc}\rangle_\Omega$.
Panel (b) shows $\langle{\rm DM}_{\rm loc}\rangle_\Omega/\langle{\rm DM}_{\rm loc}\rangle_{\Omega,\rm fid}$ over the overlapping time interval; the horizontal dashed line marks unity.
Panel (c) shows $\langle{\rm DM}_{\rm wind}\rangle_\Omega/\langle{\rm DM}_{\rm loc}\rangle_\Omega$, and panel (d) shows the normalized angular DM spread, $A_{{\rm DM},\Omega}$.
The Alpha2 model adopts $\alpha=2$ while matching the fiducial $L_{\rm w}(t_{\rm start})$, whereas the High-$M_{\rm ej}$ model increases both $M_{\rm ej}$ and $E_{\rm SN}$ by a factor of two, preserving the fiducial ejecta velocity scale.
The Alpha2 curves lie close to the fiducial results and are therefore partly obscured by the thicker fiducial curves or, in panel (b), by the unity reference line.
}
\label{fig:parameter_dependence}
\end{figure*}

We next examine how the source-local DM responds to variations in the wind parameters, ejecta properties, and simulation starting time summarized in Table~\ref{tab:model_suite}. Figure~\ref{fig:parameter_dependence} compares the solid-angle-averaged source-local DM, its ratio to the fiducial result, the wind fraction $\langle{\rm DM}_{\rm wind}\rangle_\Omega/\langle{\rm DM}_{\rm loc}\rangle_\Omega$, and the normalized angular DM spread, $A_{{\rm DM},\Omega}$. Ratios in panel (b) are evaluated over the time interval shared by each model and the fiducial run. The Smooth-bipolar model is discussed separately in Section~\ref{sec:long_term}, where its long-term evolution is compared with that of the fiducial model.

\subsection{Wind-parameter Dependence}
\label{subsec:wind_parameter_dependence}

We first consider the High-$L_0$, Narrow-cone, and Low-wind models, which increase the wind-luminosity normalization, narrow the cone opening angle, and reduce the wind velocity, respectively. The Narrow-cone model concentrates the same total wind power into a smaller polar opening angle, while the Low-wind model reduces $v_{\rm w}$ at fixed luminosity. For the fixed kinetic fraction adopted here, the corresponding mass-injection rate scales as $\dot{M}_{\rm w}\propto L_{\rm w}/v_{\rm w}^{2}$, while the boundary density scales as $\rho_{\rm w}\propto L_{\rm w}/v_{\rm w}^{3}$. The Low-wind model therefore represents a more strongly mass-loaded effective outflow.

Figures~\ref{fig:parameter_dependence}(a) and (b) show that $\langle{\rm DM}_{\rm loc}\rangle_\Omega$ decreases with time in all three wind-parameter models. Their mean DMs remain within approximately 10\% of the fiducial result over most of the simulated interval: High-$L_0$ produces a slightly lower DM, whereas Narrow-cone and Low-wind give modestly higher values. These parameter changes produce larger fractional differences in the wind contribution than in the source-local DM. As shown in Figure~\ref{fig:parameter_dependence}(c), the Low-wind model reaches $\langle{\rm DM}_{\rm wind}\rangle_\Omega/\langle{\rm DM}_{\rm loc}\rangle_\Omega\simeq0.1$ at late times, while High-$L_0$ reaches a few percent and Narrow-cone remains comparable to or below the fiducial result. Even in the more strongly mass-loaded Low-wind model, however, the wind remains a subdominant contributor to the angle-averaged electron column.

Changing the wind parameters affects the angular DM spread more strongly than the mean DM, but the effect differs among the models. Figure~\ref{fig:parameter_dependence}(d) shows that High-$L_0$ increases $A_{{\rm DM},\Omega}$ to approximately 2 at late times, compared with about 0.7 in the fiducial model. By contrast, the Narrow-cone and Low-wind models remain comparable to or below the fiducial angular spread. The modest change in the Narrow-cone model likely reflects the percentile-based definition of $A_{{\rm DM},\Omega}$. Because the enhanced anisotropy is confined to a small polar solid angle, it has limited influence on the weighted \(P_{10,\Omega}\)--\(P_{90,\Omega}\) range. Thus, the wind parameters can strongly change how DM varies with viewing angle while leaving the solid-angle-averaged DM nearly unchanged.

We also test the sensitivity to the temporal decay of the wind luminosity using the Alpha2 model. This model adopts the spin-down-like value \(\alpha=2\) and adjusts the luminosity normalization to \(L_0=8.4\times10^{42}\ \mathrm{erg\,s^{-1}}\), so that the initial wind luminosity matches the fiducial value, \(L_{\rm w}(t_{\rm start})\simeq4.5\times10^{41}\ \mathrm{erg\,s^{-1}}\). The Alpha2 model closely follows the fiducial result in all four panels of Figure~\ref{fig:parameter_dependence}. Thus, matching the initial wind luminosity while adopting a steeper temporal decay produces little change in the mean DM, wind fraction, or angular spread over the simulated interval. The electron column remains dominated by ejecta-associated nonwind material.

\subsection{Effects of Ejecta Mass and Starting Time}
\label{subsec:ejecta_start_dependence}

We next examine the effects of changing the ejecta mass and the simulation starting time. In the High-$M_{\rm ej}$ model, both the ejecta mass and the supernova kinetic energy are increased by a factor of two, to $M_{\rm ej}=6.0\,M_\odot$ and $E_{\rm SN}=2.0\times10^{51}\ {\rm erg}$. Because the characteristic ejecta velocity scales as $v_{\rm ej}\propto(E_{\rm SN}/M_{\rm ej})^{1/2}$, this choice preserves the fiducial velocity scale while increasing the ejecta density and column density. Increasing $M_{\rm ej}$ alone at fixed $E_{\rm SN}$ would lower the ejecta velocity and change the fraction of the ejecta lying outside the fixed inner boundary, making it more difficult to isolate the effect of a larger ejecta column.

Figures~\ref{fig:parameter_dependence}(a) and (b) show that the High-$M_{\rm ej}$ model increases $\langle{\rm DM}_{\rm loc}\rangle_\Omega$ by approximately a factor of two over most of the simulated interval. This increase closely follows the factor of 2 increase in the ejecta mass, while the DM continues to decrease with time at a rate similar to that in the fiducial model. The wind fraction remains small, and the angular DM spread changes less strongly than the mean DM. The ejecta mass therefore mainly changes the overall DM level, whereas the decline is still driven by the expansion and dilution of the ejecta and swept-up shell.

We also examine a model initialized at a later source age, $t_{\rm start}=2\ {\rm yr}$. Changing $t_{\rm start}$ modifies the initial ejecta structure and the wind luminosity at the beginning of the calculation. At the fixed inner radius, the ejecta velocity corresponding to the inner boundary is $v_{\rm in}=r_{\rm in}/t_{\rm start}$, so a later starting time also changes how much of the slow inner ejecta is represented in the computational domain. Figure~\ref{fig:parameter_dependence}(b) shows that this model has a mean DM approximately three to four times larger than the fiducial result over most of the overlapping time interval. The higher DM results from several changes in the initial conditions rather than from varying a single wind parameter. In particular, the later starting time allows more slow inner ejecta to remain outside the fixed inner boundary. Although the overall DM level is higher, $\langle{\rm DM}_{\rm loc}\rangle_\Omega$ still decreases with time. The wind also remains a minor contributor to the electron column.

Taken together, the parameter survey shows that the wind properties mainly regulate the wind fraction and angular structure, whereas the ejecta mass and simulation starting time produce larger changes in the mean DM normalization. None of these variations changes the main trend: the solid-angle-averaged source-local DM decreases as the system expands and remains dominated by nonwind material.

\section{Long-term Evolution and Wind-profile Robustness}
\label{sec:long_term}

\begin{figure*}
\centering
\includegraphics[width=\textwidth]{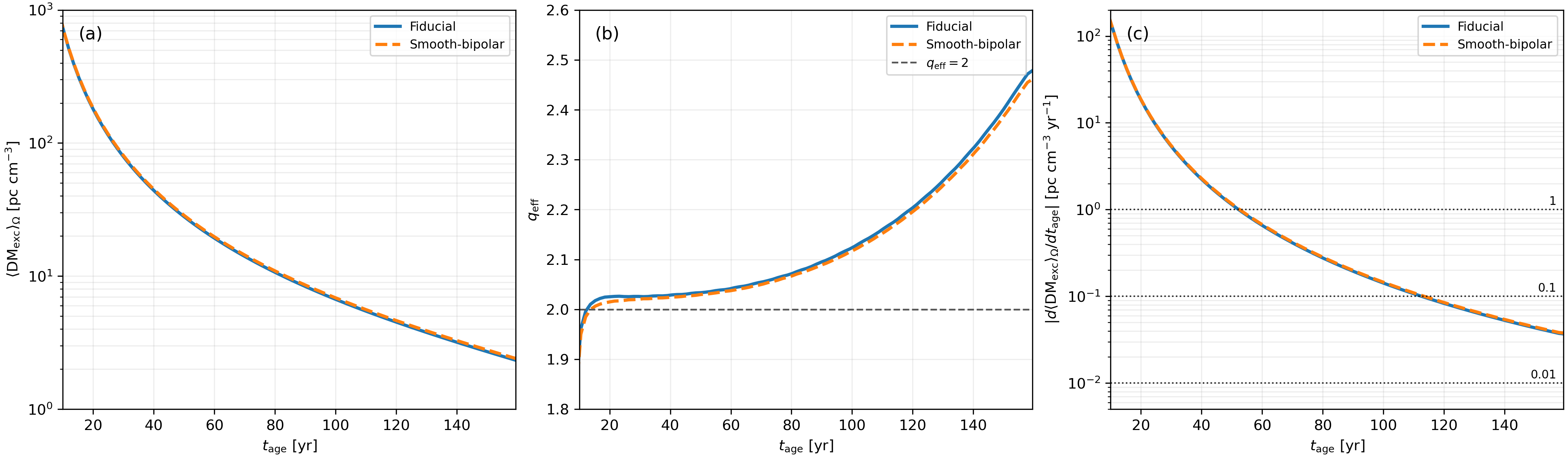}
\caption{
Long-term evolution of the ambient-subtracted excess DM for the Fiducial and Smooth-bipolar models, shown by the solid and dashed curves, respectively.
Panel (a) shows the solid-angle-averaged excess DM, $\langle{\rm DM}_{\rm exc}\rangle_\Omega$.
Panel (b) shows the effective decline index,
$q_{\rm eff}=-d\ln\langle{\rm DM}_{\rm exc}\rangle_\Omega/d\ln t_{\rm age}$; the horizontal dashed line marks $q_{\rm eff}=2$.
Panel (c) shows the absolute decline rate,
$\left|d\langle{\rm DM}_{\rm exc}\rangle_\Omega/dt_{\rm age}\right|$.
The dotted horizontal lines in panel (c) mark rates of $1$, $0.1$, and $0.01\ {\rm pc\ cm^{-3}\ yr^{-1}}$.
The close agreement between the two models shows that the long-term excess-DM evolution is only weakly affected by the detailed angular distribution of the injected wind.
}
\label{fig:long_term}
\end{figure*}

We next examine whether the declining source-local DM persists beyond the first several decades and whether it depends on the angular distribution of the injected wind. Figure~\ref{fig:long_term} compares the Fiducial and Smooth-bipolar models. In the fiducial model, the wind power is concentrated within two polar cones, whereas in the Smooth-bipolar model it varies more gradually with polar angle over a broader angular range. The Smooth-bipolar model therefore tests whether the long-term DM evolution depends on the detailed angular form of the wind injection; its angular profile is described in Appendix~\ref{app:smooth_bipolar_wind}. For this long-term comparison, we use the ambient-subtracted excess DM, ${\rm DM}_{\rm exc}$, defined in Equation~(\ref{eq:dm_excess}). During the first several decades, the ambient column is much smaller than the contribution from the ejecta and swept-up shell. At later times, however, the ejecta column decreases substantially, and the fixed ambient column across the computational domain becomes a nonnegligible fraction of the total DM. Subtracting this contribution isolates the evolving electron column associated with the source environment.

Figure~\ref{fig:long_term}(a) shows that $\langle{\rm DM}_{\rm exc}\rangle_\Omega$ continues to decrease from approximately $7\times10^{2}\ {\rm pc\ cm^{-3}}$ at $t_{\rm age}\simeq10\ {\rm yr}$ to a few ${\rm pc\ cm^{-3}}$ at $t_{\rm age}\simeq150\ {\rm yr}$. The Fiducial and Smooth-bipolar curves remain close throughout this interval. Replacing the cone-concentrated wind with a smoother angular distribution therefore produces little change in the long-term DM level.

To describe how rapidly the excess DM decreases relative to the source age, we define the effective decline index as
\begin{equation}
q_{\rm eff}
\equiv
-\frac{d\ln\langle{\rm DM}_{\rm exc}\rangle_\Omega}
{d\ln t_{\rm age}}.
\label{eq:q_eff}
\end{equation}
For \(\langle{\rm DM}_{\rm exc}\rangle_\Omega\propto t_{\rm age}^{-q}\), this definition gives \(q_{\rm eff}=q\). Figure~\ref{fig:long_term}(b) shows that \(q_{\rm eff}\) is close to 2 during the earlier evolution and gradually rises to approximately 2.5 by \(t_{\rm age}\simeq150\ \mathrm{yr}\). Thus, the excess DM evolves approximately as \(t_{\rm age}^{-2}\) during the first several tens of years and becomes modestly steeper later. The Fiducial and Smooth-bipolar models remain nearly indistinguishable in \(q_{\rm eff}\).

The gradual increase in \(q_{\rm eff}\) does not imply a larger absolute annual change because the remaining electron column becomes much smaller as the source ages. Figure~\ref{fig:long_term}(c) shows that \(\left|d\langle{\rm DM}_{\rm exc}\rangle_\Omega/dt_{\rm age}\right|\) falls below \(1\ \mathrm{pc\,cm^{-3}\,yr^{-1}}\) at an age of approximately \(60\ \mathrm{yr}\), approaches \(0.1\ \mathrm{pc\,cm^{-3}\,yr^{-1}}\) near \(10^{2}\ \mathrm{yr}\), and decreases to a few \(10^{-2}\ \mathrm{pc\,cm^{-3}\,yr^{-1}}\) by \(t_{\rm age}\simeq150\ \mathrm{yr}\). The two wind profiles again give nearly identical results. The secular decline therefore persists well beyond the early wind--ejecta interaction, but its absolute magnitude becomes progressively smaller. Its long-term behavior is controlled mainly by the expansion and dilution of ejecta-associated material rather than by the detailed angular form of the injected wind.

\section{Radial Origin of the Source-local DM}
\label{sec:radial_dm}

We now use cumulative radial DM profiles to identify where the electron column is accumulated within the wind--ejecta system and to connect the DM evolution with the wind-inflated cavity and surrounding shell. For each material component \(X\), we define the cumulative DM from the inner boundary to radius \(r\) as
\begin{equation}
{\rm DM}_{X}(<r,\theta,t_{\rm age})
=
\int_{r_{\rm in}}^{r}
n_{e,X}(r',\theta,t_{\rm age})\,dr',
\label{eq:cumulative_radial_dm}
\end{equation}
where \(X\) denotes the total, wind, or nonwind component. We then calculate the solid-angle average of each cumulative profile using the same angular weighting as in Equation~(\ref{eq:dm_angle_average}). To trace how the viewing-angle dependence accumulates with radius, we additionally calculate the solid-angle-weighted 10th and 90th percentiles, \(P_{10,\Omega}(<r)\) and \(P_{90,\Omega}(<r)\), of the cumulative total DM at each cutoff radius. Their separation provides a radial measure of the viewing-angle-dependent DM spread, using the same percentile-based characterization adopted for the full source-local DM. Unlike the long-term diagnostic in Section~\ref{sec:long_term}, we show the full cumulative DM here so that the wind and nonwind contributions add directly to the total profile. The nonwind component includes both the original ejecta and ambient gas after their hydrodynamic redistribution. The black reference curve instead shows the cumulative column that the uniform ambient medium would produce if it remained unperturbed; it is not an additional component added to the total DM.

\begin{figure*}
\centering
\includegraphics[width=\textwidth]{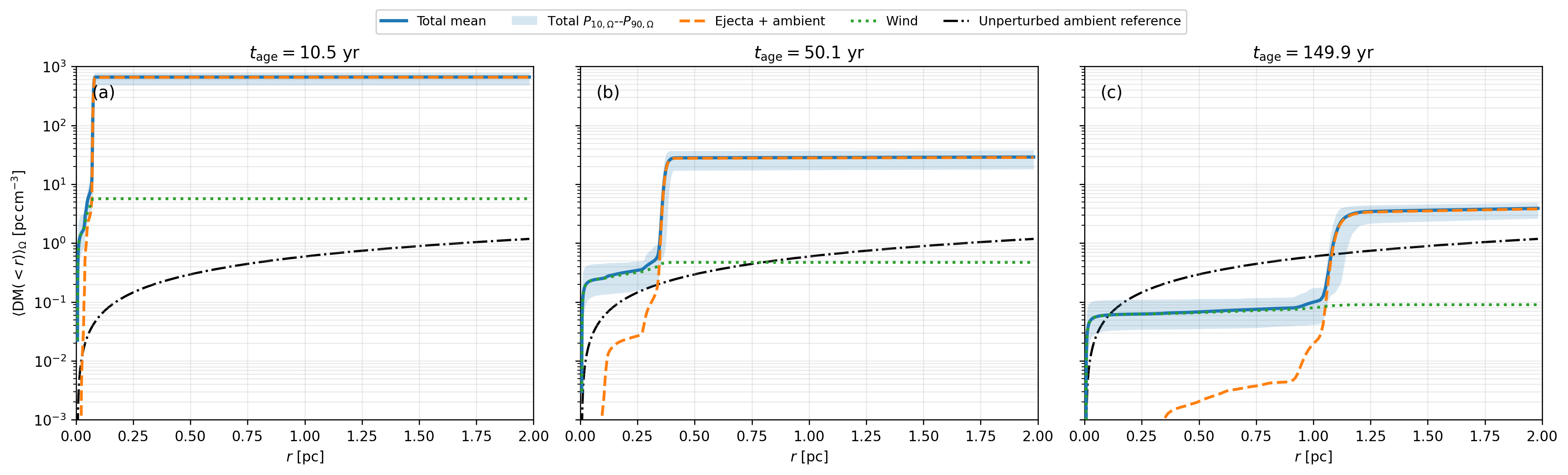}
\caption{Solid-angle-averaged cumulative radial DM profiles of the fiducial model at source ages of 10.5, 50.1, and 149.9 yr. The solid blue curve shows the mean cumulative total DM, while the blue shaded region marks the solid-angle-weighted \(P_{10,\Omega}\)--\(P_{90,\Omega}\) range of the cumulative total DM at each cutoff radius. The total DM is decomposed into the nonwind and wind contributions. The nonwind component includes the original ejecta and ambient gas after their hydrodynamic redistribution. The black dot--dashed curve shows the cumulative column of the uniform ambient medium if it remained unperturbed and is included only as a reference. At each epoch, the cumulative total DM rises sharply across the dense nonwind shell, where the angular spread also increases most strongly, while the wind contribution reaches a much lower level. Beyond the shell, the angular spread changes only modestly, indicating that most of the viewing-angle-dependent DM variation is accumulated across the compressed shell region, with relatively little additional contribution at larger radii. The shell-associated jump moves outward as the system expands.}
\label{fig:radial_dm}
\end{figure*}

Figure~\ref{fig:radial_dm} shows that the cumulative total DM increases relatively slowly through the inner wind-inflated cavity and then rises sharply over a narrow radial interval. This jump marks a dense compressed nonwind shell region in the evolving remnant. Its radius moves from approximately \(0.1\ {\rm pc}\) at 10.5 yr to approximately \(0.4\ {\rm pc}\) at 50.1 yr and \(1.1\ {\rm pc}\) at 149.9 yr. The \(P_{10,\Omega}\)--\(P_{90,\Omega}\) range remains relatively small in absolute terms through most of the cavity and broadens sharply across the same shell-associated interval. Beyond the shell, both the cumulative mean profile and the viewing-angle-dependent DM spread change only gradually with radius. This behavior indicates that most of the final viewing-angle-dependent DM variation is accumulated across the compressed shell region, with comparatively little additional angular variation contributed at larger radii.

The material decomposition shows that the wind and nonwind gas play different roles. Within the cavity, the cumulative total profile approximately follows the wind profile, particularly at the two later epochs, so the wind provides most of the small electron column accumulated there. Across the shell, however, the nonwind contribution increases rapidly and becomes the dominant component. The simultaneous increase in the cumulative DM and its viewing-angle-dependent spread indicates that most of the angular variation is associated with the compressed nonwind material in the shell region, rather than with the direct electron column of the wind itself. Outside the shell, the total and nonwind profiles nearly overlap, whereas the wind profile has already reached a much lower plateau. Thus, the wind can dominate the small column inside the cavity while contributing only a minor fraction of the total DM through the full system.

The black reference curve provides a baseline for distinguishing the column of the unperturbed ambient medium from that of the hydrodynamically redistributed nonwind gas. It rises gradually with radius because the ambient density is uniform, whereas the simulated nonwind profile rises sharply where ejecta and ambient gas have been compressed into the shell. At the earlier epochs, the unperturbed ambient column remains much smaller than the total DM. By \(t_{\rm age}\simeq150\ \mathrm{yr}\), however, it reaches approximately \(1\ \mathrm{pc\,cm^{-3}}\) across the plotted radial domain and becomes a nonnegligible fraction of the total column, which is only a few \(\mathrm{pc\,cm^{-3}}\). This behavior explains why Section~\ref{sec:long_term} uses the ambient-subtracted quantity \({\rm DM}_{\rm exc}\) to isolate the evolving source-associated column at late times. Meanwhile, the shell moves outward and the electron column accumulated across it decreases, directly linking the long-term DM decline to the dilution of shell-associated nonwind material.

\section{Comparison with Repeating FRB DM Evolution}
\label{sec:repeating_frb_comparison}

We use the long-term DM evolution of well-monitored repeaters to test whether young wind--ejecta environments can produce comparable electron columns and secular variation rates. FRB 20190520B provides the main quantitative comparison because of its rapid and approximately monotonic DM decline. FRB 20220529A represents a more slowly evolving case, whereas FRB 20121102A has a more complex, nonmonotonic history. We also compare the characteristic age and electron column obtained from our fiducial model with the one-dimensional SNR calculations of \citet{ZhangEtAl2026SNR}.

The observed DM contains contributions from the Milky Way, the intergalactic medium, intervening halos, the host galaxy, and the immediate source environment. By contrast, the simulations predict only the ambient-subtracted electron column associated with the modeled wind--ejecta system. We therefore do not match the simulated \(\langle{\rm DM}_{\rm exc}\rangle_{\Omega}\) directly to the observed total or host-associated DM. Instead, we use the secular variation rate as the principal diagnostic because the nonlocal and extended host-galaxy contributions are expected to vary negligibly over the monitoring interval. For a source at redshift \(z\), the absolute source-frame variation rate is related to the observer-frame rate by
\begin{equation}
\left|
\frac{d{\rm DM}_{\rm src}}{dt_{\rm src}}
\right|
=
(1+z)^2
\left|
\frac{d{\rm DM}_{\rm obs}}{dt_{\rm obs}}
\right|.
\label{eq:dmdt_source_frame}
\end{equation}
The quantitative comparisons below adopt the hydrogen-equivalent normalization $x_e=1$. For the approximate $q\simeq2$ evolution relevant here, adopting a lower constant electron abundance of $x_e=0.5$ would reduce the inferred rate-matching age by about 20\%. Spatially or temporally varying ionization would require a more detailed treatment.

\subsection{FRB 20190520B}
\label{subsec:frb20190520b}

FRB 20190520B is associated with a compact persistent radio source in a dwarf host galaxy at \(z=0.241\) and has a large, although uncertain, host-associated DM contribution \citep{Niu2022}. Because the fraction arising from the immediate source environment is not independently known, we match the observed secular decline rate to the fiducial model and evaluate the corresponding source-local electron column.

\begin{figure*}
\centering
\includegraphics[width=\textwidth]{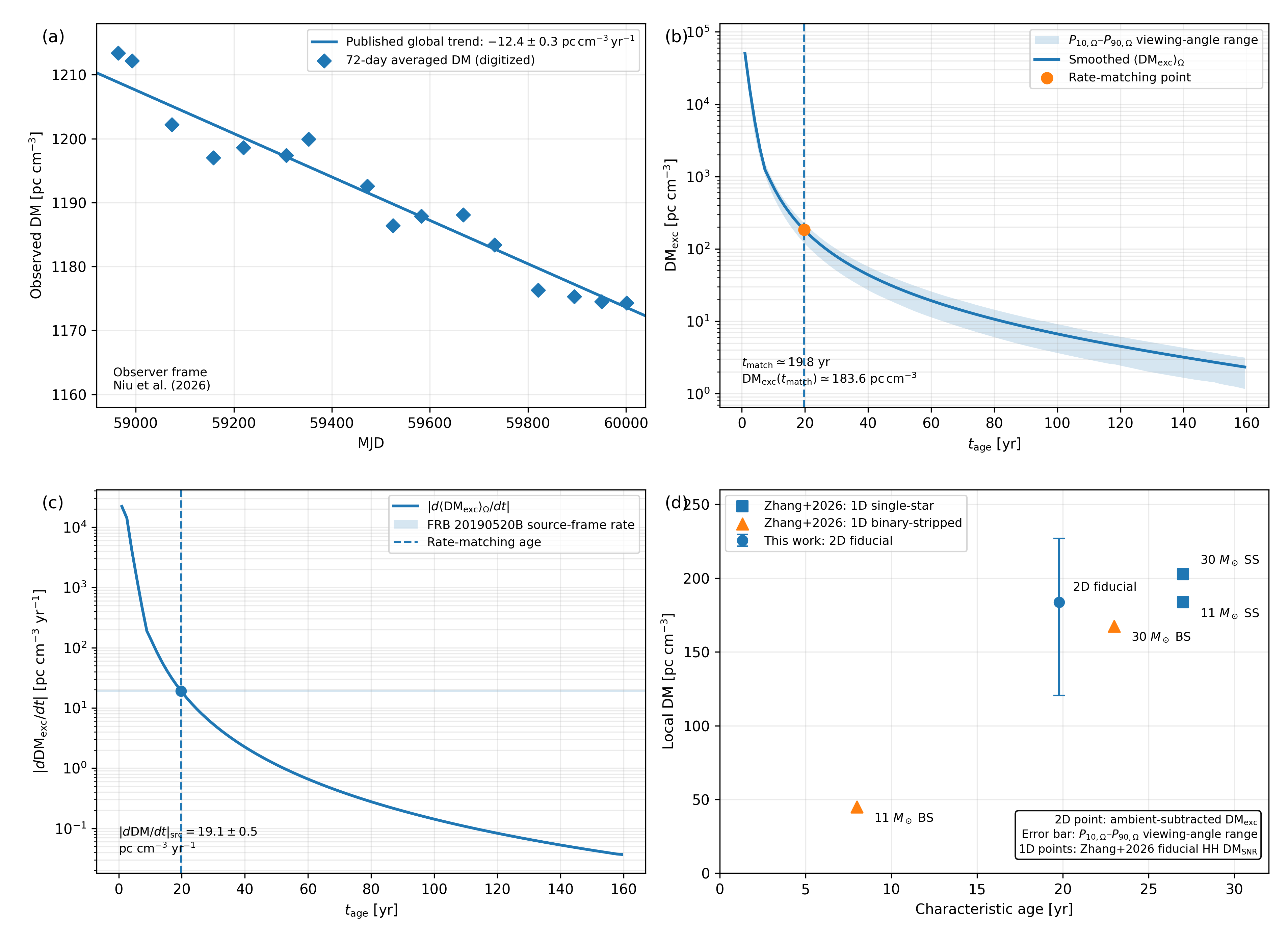}
\caption{Comparison of the fiducial simulation with the DM evolution of FRB 20190520B. Panel (a) shows the observer-frame 72 day averaged DM measurements digitized from Figure~1(a) of \citet{Niu2026}. The solid line has the published slope, \((d{\rm DM}/dt)_{\rm obs}=-12.4\pm0.3\ \mathrm{pc\,cm^{-3}\,yr^{-1}}\), with its normalization determined from the digitized averaged measurements. The digitized points are included for visualization, whereas the quantitative model comparison uses the published decline rate. Panel (b) shows the smoothed solid-angle-averaged ambient-subtracted excess DM of the fiducial model. The shaded region gives the \(P_{10,\Omega}\)--\(P_{90,\Omega}\) viewing-angle range. The vertical dashed line and marker indicate the rate-matching age, \(t_{\rm match}=19.8\ \mathrm{yr}\), at which \(\langle{\rm DM}_{\rm exc}\rangle_{\Omega}=183.6\ \mathrm{pc\,cm^{-3}}\). Panel (c) shows the absolute DM decline rate derived from the same smoothed mean-DM curve. The horizontal band gives the source-frame rate inferred for FRB 20190520B, \(19.1\pm0.5\ \mathrm{pc\,cm^{-3}\,yr^{-1}}\), and the vertical dashed line marks the corresponding rate-matching age. Panel (d) compares the characteristic age and local DM of the present two-dimensional fiducial model with those of the fiducial HH one-dimensional SNR models of \citet{ZhangEtAl2026SNR}. The error bar on the two-dimensional point gives its \(P_{10,\Omega}\)--\(P_{90,\Omega}\) viewing-angle range. Squares and triangles denote the single-star and binary-stripped progenitor models, respectively. The two studies use different DM definitions, ionization prescriptions, integration regions, dimensionalities, and frame conventions; panel (d) therefore compares characteristic age and local-column scales rather than implying strict quantitative agreement.}
\label{fig:frb20190520b_comparison}
\end{figure*}

Figure~\ref{fig:frb20190520b_comparison}(a) shows the 72 day averaged DM measurements reported by \citet{Niu2026}, digitized from their Figure~1(a). These points illustrate the long-term observational evolution; the quantitative comparison uses the decline rate reported by \citet{Niu2026} rather than a new fit to the digitized data. The values in panel (a) represent the total observed DM, including contributions from the Milky Way, the intergalactic medium and intervening halos, and the full host-galaxy column. By contrast, the simulated \(\langle{\rm DM}_{\rm exc}\rangle_{\Omega}\) represents only the ambient-subtracted electron column associated with the immediate wind--ejecta environment and should not be compared directly with the observed total DM. Over approximately four years, the total observed DM decreases at an observer-frame rate of \((d{\rm DM}/dt)_{\rm obs}=-12.4\pm0.3\ \mathrm{pc\,cm^{-3}\,yr^{-1}}\).

Using Equation~(\ref{eq:dmdt_source_frame}), this decline corresponds to an absolute source-frame rate of \(19.1\pm0.5\ \mathrm{pc\,cm^{-3}\,yr^{-1}}\). Figures~\ref{fig:frb20190520b_comparison}(b) and \ref{fig:frb20190520b_comparison}(c) show the smoothed solid-angle-averaged excess DM and its absolute time derivative, respectively. We define \(t_{\rm match}\) as the first source age at which
\(\left|d\langle{\rm DM}_{\rm exc}\rangle_{\Omega}/dt_{\rm age}\right|\)
equals the inferred source-frame rate. Under the adopted $x_e=1$ normalization, the fiducial model reaches this rate at \(t_{\rm match}\simeq19.8\ \mathrm{yr}\), when \(\langle{\rm DM}_{\rm exc}\rangle_{\Omega}\simeq183.6\ \mathrm{pc\,cm^{-3}}\). This value is consistent with those from the fits of RM and persistent radio source \citep{WFY2025,Zhao2026}.
The corresponding \(P_{10,\Omega}\)--\(P_{90,\Omega}\) viewing-angle range is approximately \(120.5\)--\(227.0\ \mathrm{pc\,cm^{-3}}\).

The rate-matching result shows that a young wind--ejecta environment can produce the observed rapid decline while retaining a source-local electron column of order \(10^{2}\ \mathrm{pc\,cm^{-3}}\). This evolving component can be superposed on larger, nearly constant foreground and host-galaxy contributions. The \(P_{10,\Omega}\)--\(P_{90,\Omega}\) range also shows that the column along an individual sight line can differ appreciably from the solid-angle average. However, the inferred age is not unique because the relation between decline rate and age depends on the ejecta properties, wind history, ionization state, and viewing direction. The comparison therefore establishes characteristic scales rather than a source-specific fit.

Figure~\ref{fig:frb20190520b_comparison}(d) compares the two-dimensional result with the fiducial high--high ionization (HH) models of \citet{ZhangEtAl2026SNR}, for which
\(\chi_{e,\mathrm{unej}}=\chi_{e,\mathrm{ISM}}=0.1\). Here ISM denotes the interstellar medium.
Using their observer-frame rate-matching convention, they obtained occurrence ages of \(27\), \(8\), \(27\), and \(23\ \mathrm{yr}\) for the \(11\,M_{\odot}\) single-star, \(11\,M_{\odot}\) binary-stripped, \(30\,M_{\odot}\) single-star, and \(30\,M_{\odot}\) binary-stripped models, respectively. At these rate-matched ages, \citet{ZhangEtAl2026SNR} obtain
\({\rm DM}_{\rm SNR}=183.6\), \(44.9\), \(202.8\), and
\(167.5\ \mathrm{pc\,cm^{-3}}\) for the same four models. Except for the \(11\,M_{\odot}\) binary-stripped case, these values overlap the characteristic age and electron-column scales obtained from our fiducial model. The lower column in the \(11\,M_{\odot}\) binary-stripped model mainly reflects its smaller ejecta mass.

This comparison remains qualitative. The \({\rm DM}_{\rm SNR}\) of \citet{ZhangEtAl2026SNR} includes unshocked ejecta, shocked plasma, and circumstellar or interstellar material within their adopted radial integration region. Our \({\rm DM}_{\rm exc}\) is defined relative to the unperturbed ambient column and includes the viewing-angle-dependent redistribution of wind, ejecta, and swept-up ambient material. The two calculations also differ in dimensionality, ionization treatment, and rate-matching convention. Despite these differences, both place the rapidly evolving component of FRB 20190520B in a young environment with an age of a few tens of years and an electron column of order \(10^{2}\ \mathrm{pc\,cm^{-3}}\).

\subsection{Additional Repeating FRB Cases}
\label{subsec:additional_repeaters}

FRB 20220529A provides a complementary example of a slower secular decline. A 3.2 yr monitoring campaign measured an observer-frame rate of
\((d{\rm DM}/dt)_{\rm obs}=-0.881\pm0.001\ \mathrm{pc\,cm^{-3}\,yr^{-1}}\)
\citep{Pandhi2026}, more than an order of magnitude smaller in absolute value than that of FRB 20190520B. Within an expansion-dominated interpretation, this slower evolution is consistent with a later or lower-column stage, although the mapping between decline rate, age, and local DM is not unique. We therefore use this source only as a qualitative comparison.

In the fiducial HH models of \citet{ZhangEtAl2026SNR}, matching the observed decline of FRB 20220529A gives ages of \(18\)--\(74\ \mathrm{yr}\) and local SNR DMs of \(12.4\)--\(36.1\ \mathrm{pc\,cm^{-3}}\), depending on the progenitor. For every progenitor considered in that study, the rate matching shifts to a later epoch and a lower electron column than for FRB 20190520B. These values illustrate the general trend toward slower DM evolution as an expanding environment ages and its electron column decreases.

FRB 20121102A has a qualitatively different history: its DM initially increased and subsequently entered a declining phase \citep{Wang2025,Snelders2025}. The late-time observer-frame decline rate is
\((d{\rm DM}/dt)_{\rm obs}=-3.93\pm0.11\ \mathrm{pc\,cm^{-3}\,yr^{-1}}\),
intermediate between the rates of FRB 20190520B and FRB 20220529A. An expanding SNR or wind--ejecta component could contribute to this late decline, but the present simulations cannot reproduce the preceding rise or the turnover. We therefore do not use FRB 20121102A as the principal quantitative comparison.

Matching only the late declining segment, the fiducial one-dimensional models of \citet{ZhangEtAl2026SNR} give characteristic ages of \(11\)--\(42\ \mathrm{yr}\) and local SNR DMs of \(26.2\)--\(97.7\ \mathrm{pc\,cm^{-3}}\). These values support the possible presence of an SNR-like declining component but do not explain the full history. The earlier increase and subsequent turnover may require an additional time-dependent plasma component, such as a companion wind or circumstellar disk in a binary system, changes in the ionization state, or coupled evolution of the SNR and central wind nebula.

The three repeaters should not be interpreted as points on a single universal evolutionary sequence because the relation between DM variation, age, and electron column depends on the source properties and viewing geometry. The present simulations are most directly applicable to the secularly declining, expansion-driven component; additional plasma or ionization evolution is required for more complex DM histories.

\section{Discussion}
\label{sec:discussion}

\subsection{Physical Origin of the DM Evolution}
\label{subsec:physical_origin_dm}

The secular decrease in the source-local DM is primarily caused by the expansion and dilution of the ejecta and wind-driven shell. For freely expanding material with an approximately fixed ionization fraction, the electron density decreases as \(n_e\propto t_{\rm age}^{-3}\), while the characteristic path length increases as \(l\propto t_{\rm age}\). The electron column therefore scales approximately as \({\rm DM}\sim n_e l\propto t_{\rm age}^{-2}\). The simulated flow is not perfectly homologous because the central wind continuously injects energy and redistributes the ejecta. Nevertheless, as shown in Figure~\ref{fig:long_term}, the excess DM initially declines approximately as \(t_{\rm age}^{-2}\). At later times, \(q_{\rm eff}\) gradually rises above 2, indicating a somewhat faster fractional decline than the \(t_{\rm age}^{-2}\) scaling. The absolute decline rate, however, continues to decrease because the electron column becomes much smaller as the source ages.

The wind and nonwind material play different roles in this evolution. As shown by the cumulative radial DM profiles in Figure~\ref{fig:radial_dm}, the wind inflates the central cavity and redistributes and compresses the surrounding gas into a shell, but the dilute wind itself supplies only a small fraction of the total electron column. Most of the source-associated DM is instead provided by the ejecta and swept-up nonwind shell. The wind is therefore dynamically important because it controls the geometry and compression of the system, whereas the denser nonwind material dominates the electron column.

Despite the strong polar concentration of the injected wind power, the swept-up outer shell remains broadly rounded and the global DM response is much less anisotropic than the wind injection itself. This behavior is consistent with lateral pressure redistribution within the wind-inflated bubble and with the dominance of ejecta-associated material in the electron column. The wind anisotropy nevertheless remains visible through the enhanced polar wind fraction and the less smooth DM evolution along the polar sight line.

The one-dimensional SNR calculations of \citet{ZhangEtAl2026SNR} provide a complementary physical decomposition of the evolving electron column. In their models, the unshocked ejecta dominate the time-varying DM, while the shocked region generally contributes only \(\lesssim 10\ \mathrm{pc\,cm^{-3}}\). In our calculations, the directly injected wind also contributes only a small fraction of the DM, although it redistributes the ejecta and strongly shapes the multidimensional morphology. Both studies therefore identify ejecta-associated material as the main evolving electron column. The one-dimensional models provide a more detailed treatment of progenitor structure, composition, ionization, and cooling, whereas the present two-dimensional simulations isolate the effects of anisotropic wind dynamics, angular redistribution, and viewing geometry.

Physical parameters and viewing geometry determine how the general expansion-driven decline appears along an individual sight line. Variations in wind power, collimation, luminosity history, and ejecta mass modify the DM normalization, angular spread, and early-time evolution. Across the model suite, however, the persistent long-term decrease is only weakly sensitive to the detailed angular form of the injected wind. Global expansion and dilution therefore control the secular decline, whereas the source parameters and viewing geometry regulate its normalization, anisotropy, and detailed temporal structure.

\subsection{Implications for Young Repeating FRBs}
\label{subsec:young_repeating_frbs}

A persistent secular decrease in DM provides a natural signature of an expanding source environment around a young repeating FRB. In the wind--ejecta picture considered here, the source-local electron column decreases as the ejecta and wind-compressed structures expand. Long-term DM monitoring can therefore reveal plasma that evolves on timescales of years to decades. A declining DM is not, however, unique to an expanding SNR environment. Changes in the ionization state, motion of compact plasma structures across the line of sight, or evolution of ionized gas elsewhere in the host galaxy can also produce time-dependent DM. The present model should therefore be regarded as a physically motivated explanation for a secularly declining component rather than as the only possible interpretation.

Recent CHIME/FRB studies further broaden the observed range of long-term DM evolution among repeating FRBs. \citet{Cook2026} reported monotonic DM variations in four repeaters on years-long timescales, while \citet{Cui2026} identified seven repeaters with significant secular DM evolution, including five decreasing and two increasing trends. The predominance of decreasing cases is qualitatively consistent with an expanding source environment such as that modeled here, whereas the increasing cases likely require additional source-local plasma evolution beyond the present fixed-ionization hydrodynamic framework.

The combination of the local DM amplitude and its variation rate is more informative than either quantity alone. If the evolving source-local component approximately follows \({\rm DM}_{\rm loc}\propto t_{\rm age}^{-q}\), its characteristic evolutionary timescale can be estimated as \(t_{\rm age}\sim q\,{\rm DM}_{\rm loc}/|d{\rm DM}_{\rm loc}/dt_{\rm age}|\). This relation explains why a large electron column accompanied by a rapid decline generally points to a young and rapidly evolving environment, whereas a smaller decline rate is more naturally associated with a later stage. It should not be interpreted as a direct age measurement. The source-local fraction of the observed DM is uncertain, \(q\) can change with time, and the inferred quantities depend on the ejecta properties, ionization state, wind history, and viewing direction.

The comparisons in Section~\ref{sec:repeating_frb_comparison} show that expansion-driven evolution can produce a range of secular DM declines, from the rapidly evolving, relatively high-column environment inferred for FRB 20190520B to the slower and lower-column case represented by FRB 20220529A. The nonmonotonic history of FRB 20121102A also demonstrates the limits of a single expansion-driven component. The present model is therefore most directly applicable to the secularly declining part of the DM evolution, whereas departures from monotonic behavior may require additional time-dependent plasma structures, binary-induced circumstellar variability, changes in the ionization state, or coupled evolution of the SNR and central wind nebula.

Viewing-angle effects introduce an additional source of diversity. A single repeater samples only one line of sight through a nonspherical environment, and its measured local column may differ appreciably from the solid-angle average. Two systems with similar ages and global properties can therefore show different DM amplitudes, while systems with different ejecta masses, ionization states, or central-engine histories need not follow the same trajectory in the DM--variation-rate plane. The simulations predict only a broad statistical tendency: younger systems should generally have larger source-local electron columns and faster secular changes, while both quantities decrease as the environment expands. Continued monitoring can test whether repeaters with larger inferred local DMs also tend to show faster declines and whether the decline rates of individual sources decrease with time. Significant departures from this behavior, particularly nonmonotonic evolution, would indicate additional plasma components or time-dependent ionization.

\subsection{Rotation Measure and Radio Transparency}
\label{subsec:rm_transparency}

The source-local plasma can affect not only the DM but also the Faraday rotation and radio transparency of a young repeating FRB environment. The present hydrodynamic simulations do not include magnetic fields or a self-consistent calculation of the ionization and thermal states. We therefore use the simulated electron column to construct illustrative RM and free--free opacity scalings rather than time-dependent predictions of these observables.

For a single dominant Faraday-rotating screen, the source-frame rotation measure can be written as \citep{Burn1966,Brentjens2005}
\nopagebreak[4]
\begingroup
\setlength{\abovedisplayskip}{4pt}
\setlength{\abovedisplayshortskip}{4pt}
\begin{equation}
{\rm RM}_{\rm src}
=
0.812
\int
\left(\frac{n_e}{\mathrm{cm}^{-3}}\right)
\left(\frac{B_{\parallel}}{\mu\mathrm{G}}\right)
\left(\frac{dl}{\mathrm{pc}}\right)
\mathrm{rad\,m^{-2}},
\label{eq:rm_definition}
\end{equation}
\endgroup
where \(B_{\parallel}\) is the line-of-sight magnetic-field component. For a source at redshift \(z\), the observed rotation measure is related to the intrinsic value by \({\rm RM}_{\rm obs}={\rm RM}_{\rm src}/(1+z)^2\). If the field does not reverse sign along the propagation path, Equation~(\ref{eq:rm_definition}) can be approximated as \({\rm RM}_{\rm src}\simeq0.812\,{\rm DM}_{\rm loc}\langle B_{\parallel}\rangle_{n_e}\), where \({\rm DM}_{\rm loc}\) is measured in \(\mathrm{pc\,cm^{-3}}\) and \(\langle B_{\parallel}\rangle_{n_e}\) is the electron-density-weighted line-of-sight field in \(\mu\mathrm{G}\).

At the rate-matching age for FRB 20190520B, \(t_{\rm match}\simeq19.8\ \mathrm{yr}\), the fiducial model has \(\langle{\rm DM}_{\rm exc}\rangle_{\Omega}\simeq183.6\ \mathrm{pc\,cm^{-3}}\). For this representative estimate, we identify the modeled source-associated column with \({\rm DM}_{\rm loc}\simeq\langle{\rm DM}_{\rm exc}\rangle_{\Omega}\). A coherent electron-density-weighted line-of-sight field of \(0.1\ \mathrm{mG}\) would then give a representative intrinsic RM of approximately \(1.5\times10^{4}\ \mathrm{rad\,m^{-2}}\), while a field of \(1\ \mathrm{mG}\) would give approximately \(1.5\times10^{5}\ \mathrm{rad\,m^{-2}}\).

These estimates show that the modeled electron column can support a large RM if the surrounding plasma contains an ordered sub-mG to mG magnetic field. Field reversals, turbulence, and a small line-of-sight projection can nevertheless substantially reduce the net RM. A useful parameterization is \(B^2/(8\pi)=\epsilon_B P\), where \(P\) is the local thermal pressure and \(\epsilon_B\) is the assumed ratio of magnetic to thermal pressure. The line-of-sight component may then be written as \(B_{\parallel}=f_{\parallel}(8\pi\epsilon_B P)^{1/2}\), where \(f_{\parallel}\) represents the field orientation and coherence. Neither \(\epsilon_B\) nor \(f_{\parallel}\) is determined by the present simulations, so a self-consistent RM calculation requires an MHD treatment.

A separate requirement is that the surrounding plasma be sufficiently transparent for GHz FRB radiation to escape. The relevant density diagnostic is the emission measure, \({\rm EM}=\int n_e^2\,dl\), expressed in \(\mathrm{pc\,cm^{-6}}\). For ionized gas near \(10^4\ \mathrm{K}\), the free--free optical depth can be approximated as \citep{Mezger1967,CondonRansom2016,Wang2020}
\nopagebreak[4]
\begingroup
\setlength{\abovedisplayskip}{4pt}
\setlength{\abovedisplayshortskip}{4pt}
\begin{equation}
\begin{aligned}
\tau_{\rm ff}\simeq{}&
3.28\times10^{-7}
\left(\frac{T_e}{10^4\ \mathrm{K}}\right)^{-1.35}
\left(\frac{\nu}{1\ \mathrm{GHz}}\right)^{-2.1} \\
&\times
\left(\frac{{\rm EM}}{\mathrm{pc\,cm^{-6}}}\right).
\end{aligned}
\label{eq:freefree_optical_depth}
\end{equation}
\endgroup
Here \(\nu\) is the frequency in the rest frame of the absorbing plasma; for a source at redshift \(z\), it is related to the observed frequency by \(\nu=(1+z)\nu_{\rm obs}\). Unlike DM, which depends linearly on \(n_e\), the free--free optical depth depends on \(n_e^2\) and is therefore particularly sensitive to thin shells, dense clumps, and unresolved small-scale structure.

We adopt \(T_e=10^4\ {\rm K}\) as a representative normalization for warm ionized absorbing gas rather than as a temperature inferred from the present hydrodynamic solution. The simulations evolve a single-fluid thermal pressure and internal energy but do not separately follow the electron temperature, electron--ion equilibration, time-dependent ionization, or radiative cooling. Therefore, the hydrodynamic thermal state cannot be directly identified with the electron temperature relevant for free--free absorption. The explicit \(T_e^{-1.35}\) dependence in Equation~(\ref{eq:freefree_optical_depth}) allows the estimated opacity to be rescaled for other assumed electron temperatures.

Using the illustrative relation \({\rm EM}\simeq C_{\rm cl}{\rm DM}_{\rm loc}^2/L\), we obtain
\begin{equation}
\begin{aligned}
\tau_{\rm ff}\simeq{}&
0.11\,C_{\rm cl}
\left(\frac{T_e}{10^4\ \mathrm{K}}\right)^{-1.35}
\left(\frac{\nu}{1\ \mathrm{GHz}}\right)^{-2.1} \\
&\times
\left(\frac{{\rm DM}_{\rm loc}}
{180\ \mathrm{pc\,cm^{-3}}}\right)^2
\left(\frac{L}{0.1\ \mathrm{pc}}\right)^{-1}.
\end{aligned}
\label{eq:freefree_dm_scaling}
\end{equation}
Here \(L\) is the effective path length and \(C_{\rm cl}\geq1\) is a clumping factor. The adopted value \(L=0.1\ \mathrm{pc}\) is illustrative and should not be interpreted as a shell thickness uniquely determined by the simulation. For \(T_e=10^4\ \mathrm{K}\), \(L=0.1\ \mathrm{pc}\), and \(C_{\rm cl}=1\), the representative electron column at \(t_{\rm match}\) gives \(\tau_{\rm ff}\simeq0.11\) at a rest-frame frequency of \(1\ \mathrm{GHz}\) and \(\tau_{\rm ff}\simeq0.05\) at \(1.4\ \mathrm{GHz}\). Under these assumptions, the environment is optically thin at GHz frequencies. The opacity can nevertheless approach or exceed unity if the electrons are concentrated in a thinner or strongly clumped shell, if the electron temperature is lower, or if the source is observed at an earlier stage with a larger electron column.

The DM, RM, and free--free optical depth probe different properties of the same environment. The DM measures the integrated electron column, the RM additionally depends on the magnetic-field strength and coherence, and the free--free opacity is especially sensitive to temperature and density inhomogeneity. Together, the above scalings show that a decades-old wind--ejecta environment can retain a substantial electron column, support a large intrinsic RM, and remain transparent to GHz radiation under plausible conditions. Quantitative predictions will require simulations that include magnetic fields, time-dependent ionization, radiative cooling, and resolved or parameterized small-scale density structure.

\subsection{Model Limitations and Future Extensions}
\label{subsec:limitations}

The main qualitative conclusions are robust across the models considered here. The wind--ejecta interaction produces an expanding cavity and compressed nonwind material, the source-local electron column decreases with time, and the directly injected wind remains subdominant in the total DM. These trends persist throughout the parameter survey and in the comparison between the bipolar-cone and smooth-bipolar wind profiles. By contrast, the DM normalization, detailed decline rate, angular spread, and characteristic age inferred for an individual source remain sensitive to the adopted ejecta, wind, and ionization parameters.

The ejecta, ambient medium, and plasma microphysics are treated in simplified forms. We adopt uniform-density homologous ejecta expanding into a uniform ambient medium, whereas realistic remnants may contain radial density and composition gradients and interact with structured circumstellar material. The calculations begin at \(t_{\rm start}=1\ \mathrm{yr}\), and the finite inner boundary excludes the innermost slow ejecta from the computational domain and DM integral. In addition, the electron abundance is prescribed rather than evolved self-consistently and does not account for spatial or temporal variations arising from composition-dependent electron abundances and ionization states, while radiative cooling is neglected. Photoionization, collisional ionization, recombination, and cooling could alter both the electron column and the compression of shocked material. The one-dimensional calculations of \citet{ZhangEtAl2026SNR} include nonequilibrium ionization and radiative cooling in the shocked plasma, but the ionization of the unshocked material remains parameterized.

The simulations are two-dimensional and axisymmetric. Three-dimensional turbulence, shell fragmentation, and clumping may broaden the distribution of DM among sight lines and produce short-timescale variations that are not captured here. The injected wind should also be interpreted as an effective large-scale, mass-loaded, and partially thermalized outflow rather than the pristine relativistic wind launched near the central engine. Accordingly, our conclusion that the directly injected wind makes only a subdominant contribution to the total DM applies to the effective wind prescription explored here. A pristine relativistic $e^\pm$ wind can have a lepton column that depends on its pair multiplicity, Lorentz factor, and dissipation history \citep{CaoEtAl2017,DaiEtAl2017}, and its DM contribution is not quantitatively constrained by the present nonrelativistic hydrodynamic calculations. The Alpha2 and smooth-bipolar models test part of the uncertainty in the large-scale luminosity history and angular structure but do not span the full range of possible engine-driven outflows. Magnetic fields are also neglected, so the RM estimates in Section~\ref{subsec:rm_transparency} depend on assumed field strengths and coherence. Three-dimensional MHD calculations will be required to determine how magnetic pressure, field geometry, and plasma instabilities affect the morphology and propagation observables.

The observational interpretation is further limited by the uncertain separation of the measured DM into foreground, extended host-galaxy, and immediate source contributions. Our simulations predict the ambient-subtracted excess DM of the modeled environment, whereas observations measure the total column along a single sight line. Comparisons with \citet{ZhangEtAl2026SNR} are also affected by differences in DM definition, ionization treatment, radial integration region, dimensionality, and reference-frame convention. The results in Section~\ref{sec:repeating_frb_comparison} should therefore be interpreted as comparisons of characteristic electron-column and secular variation scales rather than as precise parameter fits. Future calculations combining realistic ejecta and circumstellar structures with time-dependent ionization, radiative cooling, and three-dimensional MHD dynamics will be needed for quantitative source modeling.

\section{Summary and Conclusions}
\label{sec:summary_conclusions}

We have used two-dimensional axisymmetric hydrodynamic simulations to investigate the interaction between a continuous anisotropic wind from a young compact object and expanding supernova ejecta. The wind inflates a low-density cavity and redistributes the ejecta into a compressed shell. Despite the strong polar concentration of the injected wind power, the swept-up outer shell remains broadly rounded, and the total electron column is much less anisotropic than the wind injection itself. The polar sight line nevertheless retains the clearest wind signature through an enhanced wind fraction and a less smooth DM evolution. The injected wind therefore strongly shapes the morphology and viewing-angle dependence but contributes only a small fraction of the source-local DM. The electron column remains dominated by the ejecta and swept-up nonwind material.

The solid-angle-averaged ambient-subtracted excess DM decreases throughout the simulated evolution. It approximately follows a \(t_{\rm age}^{-2}\) decline during the first several tens of years and becomes modestly steeper later, while the absolute DM variation rate continues to decrease. Variations in the wind and ejecta parameters modify the DM normalization, early-time evolution, angular spread, and wind fraction, but the solid-angle-averaged DM declines in all models. Increasing the ejecta column raises the DM normalization, whereas changing the wind properties primarily alters the morphology and early evolution. The similar long-term behavior of the bipolar-cone and smooth-bipolar models further shows that the mean DM decline is only weakly sensitive to the detailed angular form of the injected wind.

FRB 20190520B provides the main observational comparison \citep{Niu2026}. After converting its observed decline rate to the source frame, the fiducial model reaches a comparable absolute variation rate at an age of approximately \(20\ \mathrm{yr}\). At this epoch, the solid-angle-averaged excess DM is approximately \(1.8\times10^{2}\ \mathrm{pc\,cm^{-3}}\), with a \(P_{10,\Omega}\)--\(P_{90,\Omega}\) viewing-angle range of approximately \((1.2\)--\(2.3)\times10^{2}\ \mathrm{pc\,cm^{-3}}\). A young expanding environment can therefore retain a substantial local electron column while producing a rapid secular decrease. Except for their \(11\,M_{\odot}\) binary-stripped model, the fiducial SNR calculations of \citet{ZhangEtAl2026SNR} give broadly overlapping characteristic age and electron-column scales. Although the two studies use different dimensionalities, ionization treatments, and DM definitions, both identify ejecta-associated material as the dominant evolving electron column. The slower decline of FRB 20220529A is qualitatively consistent with a more slowly evolving, lower-column environment \citep{Pandhi2026}. By contrast, the nonmonotonic DM history of FRB 20121102A cannot be explained by a single expansion-driven component. Its late decline may contain an SNR-like contribution, whereas its earlier rise and turnover require additional time-dependent plasma structures, such as binary-induced circumstellar variability, changes in the ionization state, or coupled evolution of the SNR and central wind nebula \citep{WFY2025,Snelders2025}.

Representative RM and free--free opacity scalings indicate that an electron column of approximately \(180\ \mathrm{pc\,cm^{-3}}\) can support an intrinsic RM of order \(10^{4}\)--\(10^{5}\ \mathrm{rad\,m^{-2}}\) and remain optically thin at GHz frequencies under plausible assumptions for the magnetic field, gas temperature, effective path length, and clumping. These estimates are illustrative rather than direct predictions of the hydrodynamic model. Quantitative joint predictions of DM, RM, and radio transparency will require more realistic ejecta and circumstellar structures, time-dependent ionization, radiative cooling, and three-dimensional MHD dynamics. The present simulations nevertheless show that the expansion and dilution of ejecta-associated material in a young SNR environment can drive a long-term decline in the source-local DM, while anisotropic central-wind dynamics primarily control the morphology and detailed sight line dependence.

\begingroup
\let\internallinenumbers\relax
\let\endinternallinenumbers\relax

\begin{acknowledgments}
This work was supported by the Xiamen Natural Science Foundation
(grant No.~3502Z202571062), the High-level Talent Project of
Xiamen University of Technology (grant No.~YKJ24005R), and the
National Natural Science Foundation of China
(grant Nos.~12494575, 12273009, and 12625304).

We thank the anonymous referee for the constructive comments and suggestions that helped improve the manuscript.

The authors used OpenAI ChatGPT \citep{OpenAI2026} to assist with
language editing and manuscript organization. All scientific content,
interpretations, and conclusions were reviewed and approved by the
authors.
\end{acknowledgments}

\endgroup

\appendix
\renewcommand{\theHequation}{\Alph{section}.\arabic{equation}}

\section{Alternative Smooth-bipolar Wind Profile}
\label{app:smooth_bipolar_wind}

The fiducial model adopts the smoothed bipolar-cone wind profile described in Section~\ref{subsec:wind_injection}. To test whether the long-term source-local DM evolution depends sensitively on the detailed shape of the wind angular distribution, we perform an additional calculation with a smooth-bipolar profile. This model represents the same general physical picture of a bipolar engine-driven outflow, but replaces the conelike transition of the fiducial prescription with a continuous angular dependence.

The angular dependence of the injected wind power is written as
\begin{equation}
A_{\rm sm}(\theta)
=
\frac{1+\epsilon_{\rm bip}\cos^2\theta}
{1+\epsilon_{\rm bip}/3},
\label{eq:smooth_bipolar_profile}
\end{equation}
where \(\epsilon_{\rm bip}\) controls the polar enhancement of the wind. The denominator follows from \(\langle\cos^2\theta\rangle_\Omega=1/3\), so that
\begin{equation}
\frac{1}{4\pi}
\int A_{\rm sm}(\theta)\,d\Omega
=
\frac{1}{2}
\int_0^\pi
A_{\rm sm}(\theta)\sin\theta\,d\theta
=
1.
\label{eq:smooth_bipolar_normalization}
\end{equation}
The total injected luminosity is therefore unchanged, and the angular profile only redistributes the wind power over solid angle.

We adopt \(\epsilon_{\rm bip}=2\), giving
\begin{equation}
A_{\rm sm}(\theta)
=
\frac{3}{5}
\left(1+2\cos^2\theta\right).
\label{eq:smooth_bipolar_adopted}
\end{equation}
The corresponding angular weights are \(A_{\rm sm}=1.8\) along the polar axis and \(A_{\rm sm}=0.6\) at the equator. The polar-to-equatorial injection contrast is therefore 3.

At the inner radial boundary, the wind velocity remains independent of polar angle, while the injected density and pressure scale in proportion to \(A_{\rm sm}(\theta)\), following Equations~(\ref{eq:wind_flux}) and (\ref{eq:wind_boundary_fluxes}). The wind luminosity history, velocity, energy partition (\(\eta_{\rm kin}=0.7\) and \(\eta_{\rm th}=0.3\)), ejecta properties, ambient medium, and numerical setup are otherwise identical to those of the fiducial model. The Smooth-bipolar run therefore provides a controlled test in which only the angular form of the wind injection is changed.

As shown in Figure~\ref{fig:long_term}, the fiducial smoothed bipolar-cone model and the Smooth-bipolar model produce nearly identical long-term excess-DM amplitudes, effective decline indices, and absolute decline rates. The long-term excess-DM evolution is therefore only weakly affected by the detailed angular prescription of the injected wind and is controlled primarily by the expansion and dilution of the ejecta and swept-up nonwind shell.

\bibliographystyle{aasjournal}
\bibliography{references}

\end{document}